\documentclass[epj]{svjour}
\usepackage{latexsym}
\usepackage{graphics}
\usepackage{epsfig}
\usepackage[latin1]{inputenc}
\usepackage{rotating}
\begin{document}
\renewcommand{\textfraction}{0.00000000001}
\renewcommand{\floatpagefraction}{1.0}
\title{Photoproduction of {\boldmath{$\eta\pi$}} pairs off nucleons and deuterons}
\author{
  A.~K{\"a}ser\inst{1},
  F.~M{\"u}ller\inst{1},
  J.~Ahrens\inst{2},
  J.R.M.~Annand\inst{3},
  H.J.~Arends\inst{2},
  K.~Bantawa\inst{4},  
  P.A.~Bartolome\inst{2}, 
  R.~Beck\inst{2,5},
  A.~Braghieri\inst{6},
  W.J.~Briscoe\inst{7},
  S.~Cherepnya\inst{8},
  S.~Costanza\inst{6},
  M.~Dieterle\inst{1}, 
  E.J.~Downie\inst{2,3,7},
  P.~Drexler\inst{9},
  L.V.~Fil'kov\inst{8},
  A.~Fix\inst{10},
  S.~Garni\inst{1},
  D.I.~Glazier\inst{3,11},
  D.~Hamilton\inst{3},
  D.~Hornidge\inst{12},
  D.~Howdle\inst{3},
  G.M.~Huber\inst{13},
  I.~Jaegle\inst{1},
  T.C.~Jude\inst{11},  
  V.L.~Kashevarov\inst{2,8},
  I. Keshelashvili\inst{1}\thanks{present address: Institut f\"ur Kernphysik, Forschungszentrum J\"ulich,
  52425 J\"ulich, Germany}, 
  R.~Kondratiev\inst{14},
  M.~Korolija\inst{15},
  B.~Krusche\inst{1},
  V.~Lisin\inst{14},
  K.~Livingston\inst{3},
  I.J.D.~MacGregor\inst{3},
  Y.~Maghrbi\inst{1},
  J.~Mancell\inst{3},   
  D.M.~Manley\inst{4},
  Z.~Marinides\inst{7},  
  J.C.~McGeorge\inst{3},
  E.~McNicoll\inst{3},  
  D.~Mekterovic\inst{15},
  V.~Metag\inst{9},
  S.~Micanovic\inst{15},  
  D.G.~Middleton\inst{12},
  A.~Mushkarenkov\inst{6},  
  A.~Nikolaev\inst{2,5},
  R.~Novotny\inst{9},
  M.~Oberle\inst{1},  
  M.~Ostrick\inst{2},
  P.~Otte\inst{2},
  B.~Oussena\inst{2,7},   
  P.~Pedroni\inst{6},
  F.~Pheron\inst{1},
  A.~Polonski\inst{14},
  S.~Prakhov\inst{16},  
  J.~Robinson\inst{3},
  T.~Rostomyan\inst{1},
  S.~Schumann\inst{2,5},
  M.H.~Sikora\inst{11},  
  D.~Sober\inst{17},
  A.~Starostin\inst{16},
  Th.~Strub\inst{1},  
  I.~Supek\inst{15},
  M.~Thiel\inst{9},
  A.~Thomas\inst{2},
  M.~Unverzagt\inst{2,5},
  N.K.~Walford\inst{1},
  D.P.~Watts\inst{11},
  D.~Werthm\"uller\inst{1,3},
  L.~Witthauer\inst{1} 
\newline(The A2 Collaboration)
\mail{B. Krusche, Klingelbergstrasse 82, CH-4056 Basel, Switzerland,
\email{Bernd.Krusche@unibas.ch}}
}
\institute{Department of Physics, University of Basel, CH-4056 Basel, Switzerland
  \and Institut f\"ur Kernphysik, University of Mainz, D-55099 Mainz, Germany
  \and SUPA School of Physics and Astronomy, University of Glasgow, G12 8QQ, UK
  \and Kent State University, Kent, Ohio 44242, USA  
  \and Helmholtz-Institut f\"ur Strahlen- und Kernphysik, University Bonn, D-53115 Bonn, Germany
  \and INFN Sezione di Pavia, I-27100 Pavia, Pavia, Italy
  \and Center for Nuclear Studies, The George Washington University, Washington, DC 20052, USA  
  \and Lebedev Physical Institute, RU-119991 Moscow, Russia
  \and II. Physikalisches Institut, University of Giessen, D-35392 Giessen, Germany  
  \and Laboratory of Mathematical Physics, Tomsk Polytechnic University, Tomsk, Russia
  \and SUPA School of Physics, University of Edinburgh, Edinburgh EH9 3JZ, UK
  \and Mount Allison University, Sackville, New Brunswick E4L3B5, Canada
  \and University of Regina, Regina, SK S4S-0A2 Canada    
  \and Institute for Nuclear Research, RU-125047 Moscow, Russia
  \and Rudjer Boskovic Institute, HR-10000 Zagreb, Croatia 
  \and University of California Los Angeles, Los Angeles, California 90095-1547, USA
  \and The Catholic University of America, Washington, DC 20064, USA
}
\authorrunning{A. K{\"a}ser et al.}
\titlerunning{Quasi-free photoproduction of $\eta\pi$-pairs }

\abstract{Quasi-free photoproduction of $\pi\eta$-pairs has been investigated
from threshold up to incident photon energies of 1.4~GeV, respectively up 
to photon-nucleon invariant masses up to 1.9~GeV. 
Total cross sections, angular distributions, invariant-mass distributions 
of the $\pi\eta$ and meson-nucleon pairs, and beam-helicity asymmetries have
been measured for the reactions $\gamma p\rightarrow p\pi^0\eta$, 
$\gamma n\rightarrow n\pi^0\eta$, $\gamma p\rightarrow n\pi^+\eta$,
and $\gamma n\rightarrow p\pi^-\eta$ from nucleons bound inside the deuteron.
For the $\gamma p$ initial state data for free protons have also been analyzed. 
Finally, the total cross sections for quasi-free production of $\pi^0\eta$ pairs 
from nucleons bound in $^3$He nuclei have been investigated in view of
final state interaction (FSI) effects. The experiments were performed at the tagged 
photon beam facility of the Mainz MAMI accelerator using an almost $4\pi$ covering 
electromagnetic calorimeter composed of the Crystal Ball and TAPS detectors. 
The shapes of all differential cross section data and the asymmetries are very 
similar for protons and neutrons and agree with the conjecture that the reactions 
are dominated by the sequential 
$\Delta^{\star}3/2^-\rightarrow\eta\Delta(1232)\rightarrow\pi\eta N$ decay chain,
mainly with $\Delta(1700)3/2^-$ and $\Delta(1940)3/2^-$. The ratios of the magnitude 
of the total cross sections also agree with this assumption. However, the absolute 
magnitudes of the cross sections are reduced by FSI effects with respect to free 
proton data.
} 
\maketitle

\clearpage
\section{Introduction}

Photoproduction of meson pairs becomes increasingly important for the study
of the electromagnetic excitation spectrum of the nucleon. The reason is simple:
so far, many states predicted by quark models have not been observed in experiment.
But if one compares the history in nuclear spectroscopy with the situation in hadron
spectroscopy, this is not at all surprising. Many features
in nuclear structure physics (e.g. rotational and vibrational bands and the like)
had not been discovered when only decays of excited states to the nuclear
ground states had been investigated. However, this was state of the art in hadron
physics until recently. Most experimental efforts were directed to single meson
photoproduction ($\pi$, $\eta$, $\eta '$, $\omega$, $\rho$, $\Phi$,...), which 
corresponds to excited state - ground-state transitions. Production of meson
pairs is an important step to more complicated decay mechanisms involving at least 
one intermediate excited state of the nucleon. Such sequential decays are more 
probable for higher lying states for which reaction phase space no longer 
suppresses decays to the $\Delta$ resonance or to the second resonance region
with respect to ground-state decays. This is not the only reason for their
importance; one must also consider the hadron structure aspects. In the quark model,
high lying states may have both possible oscillator modes excited. For such states,
it is a reasonable conjecture that they tend to decay in a two-step process via an 
intermediate state. The intermediate state could be selected such that in the first 
transition only one oscillator mode is de-excited followed by a ground-state transition, 
which de-excites the second one. It would be difficult to identify states with such 
decay patterns in single meson production reactions and this could suppress entire 
multiplets of states in the experimental data base.

The study of multiple-meson final states is challenging. The reaction amplitudes 
for photoproduction of single pseudo-scalar mesons can be fixed by the 
measurement of at least eight carefully chosen observables \cite{Chiang_97} as a 
function of two independent kinematic variables. However, for pseudo-scalar meson 
pairs \cite{Roberts_05} the determination of the magnitude of the amplitudes already
requires the measurement of eight observables as a function of five kinematic parameters.
An extraction of the phases involves the measurement of at least 15 observables.
Therefore, `complete experiments', which are currently being discussed for single
meson production, are unrealistic. Nevertheless, recent experimental progress is 
encouraging. The systematic investigation of multiple-meson final states became possible 
due to the almost $4\pi$ solid-angle coverage of modern detector systems. In particular, 
large-angle electromagnetic calorimeters, which can identify recoil nucleons, charged pions,
and photons from the decays of neutral mesons, gave a large boost to this program. 

The best studied multiple-meson final state is the production of pion pairs, in particular 
$\pi^0$ pairs. Reactions with charged mesons are more affected by non-resonant
production processes because the photons can directly couple to the charge
of the mesons. Nevertheless, such reactions must also be studied in order to
reveal the isospin structure of the excitations. Recently, many new precise
experimental results accompanied by detailed reaction analyses became available
for pion pairs 
\cite{Sarantsev_08,Thoma_08,Krambrich_09,Kashevarov_12,Zehr_12,Oberle_13,Oberle_14,Thiel_15,Sokhoyan_15a,Sokhoyan_15b,Dieterle_15}.  

The $\eta\pi$ final state has also attracted interest. Total cross 
sections, invariant mass distributions, and some polarization observables, have 
been measured for the production of $\eta\pi^0$ pairs off protons at LNS in Sendai, 
Japan \cite{Nakabayashi_06}, GRAAL at ESRF in Grenoble, France \cite{Ajaka_08}, 
ELSA in Bonn, Germany \cite{Horn_08a,Horn_08b,Gutz_08,Gutz_10,Gutz_14}, and at 
MAMI in Mainz, Germany \cite{Kashevarov_09,Kashevarov_10} (see \cite{Krusche_15} 
for a recent summary). In comparison to pion pairs, this channel has more selectivity. 
As far as decays of nucleon resonances are concerned the iso-scalar $\eta$ 
meson can only be emitted in transitions between two $N^{(\star )}$ or between two 
$\Delta^{(\star )}$ states (but not in $N^{(\star )}\leftrightarrow \Delta^{(\star )}$ 
transitions). The $\Delta$-like resonances decay into $N\eta\pi$ mainly via two sequences, 
$\Delta^{\star}\rightarrow\eta\Delta(1232)3/2^+$ and 
$\Delta^{\star}\rightarrow\pi N(1535)1/2^-$,
whereas the $N^{\star}$ states produce the final $N\pi\eta$ state
only by pion emission to the $\pi N(1535)1/2^-$ channel. Therefore, the reaction is in 
particular sensitive to excited resonances, whose decay via the first sequence 
$\Delta^{\star}\rightarrow\eta\Delta(1232)3/2^+$ leaves its trace in the $\pi N$ invariant 
mass spectra, peaking at $\Delta$(1232). The 
$\Delta^{\star}(N^{\star})\rightarrow\pi N(1535)1/2^-\rightarrow N\pi\eta$
transitions produce $\eta$-nucleon invariant masses $m(\eta,N)$ characteristic for
the $N(1535)1/2^-$ state (i.e. close to the kinematical lower limit of 
$m(\eta,N)$=1485~MeV). At sufficiently large incident photon energies, contributions from 
the $a_0$(980) meson are expected. Its decay to $\eta\pi$ results in a peak in
the $\eta$-pion invariant mass spectrum \cite{Horn_08a,Horn_08b,Gutz_14}.
 
The analysis of the available data for the $\gamma p\rightarrow p\pi^0\eta$ reaction,
including invariant mass distributions, angular distributions, and polarization observables 
measured with circularly and linearly polarized photon beams 
\cite{Horn_08b,Gutz_14,Kashevarov_09,Kashevarov_10,Fix_10}, has revealed a strong contribution 
of the $\Delta\mbox{(1700)}3/2^-\rightarrow \eta\Delta\mbox{(1232)}\rightarrow \eta\pi^0 p$
decay chain in the threshold region. In \cite{Kashevarov_09}, the authors have shown that
with the contribution of the $\Delta\mbox{(1700)}3/2^-$ alone (no background terms, no further 
nucleon resonances), most features of
the total cross section and several types of angular distributions can be described.
With the availability of experimental results for polarization observables (beam-helicity
asymmetry in \cite{Kashevarov_10} and beam asymmetry in \cite{Ajaka_08,Gutz_08}), the model
was extended in \cite{Fix_10} to contributions from several $I=3/2$ $\Delta$-resonances and 
background terms (Born-terms). Predictions for further polarization observables from the 
extended model have been made in \cite{Fix_11}. Using data from the CBELSA/TAPS experiment,
which cover a larger energy range and provide polarization observables measured 
with linearly polarized photon beams, the reaction has also been analyzed in the framework of
the Bonn-Gatchina coupled channel model \cite{Gutz_14}. So far, all results are consistent 
with a dominant contribution from the $\Delta\mbox{(1700})3/2^-$ resonance in the threshold region. 
Therefore, this reaction is promising for a detailed investigation of this state, which is  
interesting because its nature is not yet well established. Besides the interpretation as a standard 
three-constituent-quark state, a dynamical generation within coupled-channel chiral 
unitary theory for meson-baryon scattering has also been discussed \cite{Doering_06a,Doering_06b}. 

So far, data are only available for the $\gamma p\rightarrow p\pi^0\eta$ reaction. However,
the isospin dependence of the production of $\pi\eta$ pairs is also of great interest. 
Since the $\eta$-meson is isoscalar, the isospin structure of the $\pi\eta$ photoproduction 
amplitude is identical to that for single $\pi$ photoproduction. Its different charge 
channels may be represented as:
\begin{eqnarray}
\label{eq:iso}
A(\gamma p\rightarrow n\pi^+\eta) & = &
-\sqrt{\frac{1}{3}}\;A^{V3}+\sqrt{\frac{2}{3}}(A^{IV}-A^{IS}) \\
A(\gamma p\rightarrow p\pi^o\eta) & = &
+\sqrt{\frac{2}{3}}\;A^{V3}+\sqrt{\frac{1}{3}}(A^{IV}-A^{IS})\nonumber\\
A(\gamma n\rightarrow p\pi^-\eta) & = &
+\sqrt{\frac{1}{3}}\;A^{V3}-\sqrt{\frac{2}{3}}(A^{IV}+A^{IS})\nonumber\\
A(\gamma n\rightarrow n\pi^o\eta) & = &
+\sqrt{\frac{2}{3}}\;A^{V3}+\sqrt{\frac{1}{3}}(A^{IV}+A^{IS})\;.\nonumber
\end{eqnarray}
where $A^{IS}$ is the isoscalar matrix element, $A^{IV}$ is the isovector, and $A^{V3}$ is 
the isospin changing one. Only the latter contributes to the excitation of $\Delta$ states.

From the decomposition of Eq.~\ref{eq:iso}, one easily sees that for the reaction chain
$\gamma N\rightarrow\Delta^{\star}\rightarrow N\pi\eta$ ($A^{IS},A^{IV}=0$),
the cross section ratios for production of neutral and charged pions off protons and
neutrons are related by: 
\begin{eqnarray} 
 \sigma (\gamma p\rightarrow \eta\pi^0 p) & = &
 \sigma (\gamma n\rightarrow \eta\pi^0 n) =\nonumber\\
2\sigma (\gamma p\rightarrow \eta\pi^+ n) & = &
2\sigma (\gamma n\rightarrow \eta\pi^- p)\;.
\label{eq:isorel}
\end{eqnarray}
Any deviations from this relation would signal contributions from $I=1/2$ $N^{\star}$
resonances or from non-resonant background. In the present work, all four isospin 
channels have been studied using liquid hydrogen and liquid deuterium 
targets. The total cross sections and their ratios already have been published in a 
preceding letter \cite{Kaeser_15}. The present paper summarizes the results for 
differential cross sections (invariant mass distributions of the meson - meson and 
meson - nucleon pairs and angular distributions), as well as the results for the helicity 
asymmetry $I^{\odot}$, which was measured with a circularly polarized photon beam and an 
unpolarized target.

\section{Definition of observables}
\label{sec:observables}

All differential cross sections presented in this paper were normalized to the respective
total cross sections in order to facilitate the comparison of different reaction channels
and between experimental results and model predictions. A kinematic reconstruction of the 
final state (as described in \cite{Krusche_11}) was used to determine the photon - nucleon 
center-of-momentum (cm) energy $W$, the kinetic energy of the recoil nucleon, and all relevant 
cm angles (see Sec.~\ref{sec:ana}). It uses as input the incident photon energy and the 
measured four momenta of the pions and $\eta$ mesons and the polar and azimuthal angles 
of the detected recoil nucleons. The effects from nuclear Fermi motion were thus removed. 
Due to the small total cross sections (maximum values around 2.5$~\mu b$ for $\pi^0$ mesons 
and around 1.5~$\mu b$ for charged mesons), the statistical quality of the data was only
sufficient to separate the total range of photon - nucleon invariant mass from 
$W$=1.7 - 1.9~GeV into four bins.  

The first group of experimental data comprises the invariant mass distributions of 
meson - meson and meson - nucleon pairs. Their definition as the magnitude of the 
sum of the four momenta of the particle pairs is straight-forward. Previous results are only 
available for the $\pi^0\eta$ final state for free protons (see \cite{Gutz_14} and Refs. 
therein).

\begin{figure}[htb]
\centerline{\resizebox{0.50\textwidth}{!}{%
  \includegraphics{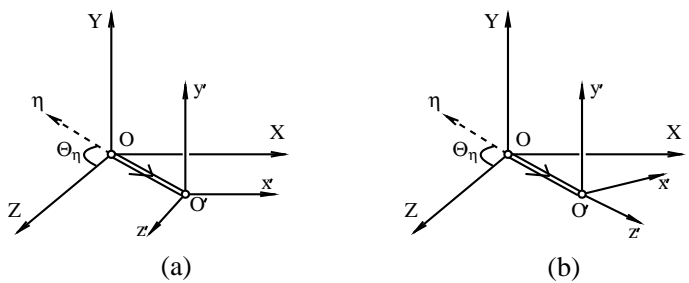}
}}
\caption{Definition of coordinate frames for angular distributions in the nucleon - pion
rest frames for (a) the canonical system, and (b) the helicity system \cite{Kashevarov_09}. 
In the canonical system (a) the $z'$ axis is parallel to the photon beam direction, in the 
helicity system (b), it is in direction of the combined $\pi N$ momentum (thick arrow). 
The $x'$ axis is in the reaction plane in both systems and the $y'$ axis is 
perpendicular to this plane.
}
\label{fig:angles}       
\end{figure}

Angular distributions have been analyzed in the same way as in \cite{Kashevarov_09,Fix_10}, 
which is adapted to the hypothesis of a dominant 
$\Delta^{\star}\rightarrow\Delta(1232)\eta\rightarrow\eta\pi N$ decay. The axes of the
photon - nucleon cm system are denoted by $X$, $Y$, $Z$ (the photon is in the direction of the $Z$ axis 
and the incident nucleon is in the $-Z$ direction). Note that the $Z$ direction is not necessarily 
parallel to the laboratory beam axis because the incident nucleon may have a momentum in a 
different direction due to Fermi motion. The momenta of the $\eta$ meson and the combined $\pi N$ 
system are back-to-back in this overall cm frame. Angular distributions of the pions in the 
pion - nucleon cm system have been extracted in two different frames, as shown in 
Fig.~\ref{fig:angles}. In the canonical frame, the $z'$-axis is parallel to the photon 
direction, while in the helicity frame, it is in direction of the momentum of the combined 
$N\pi$ system, i.e. in the direction of the momentum vector of the supposed intermediate 
$\Delta(1232)$ resonance. The direction of the $y'$ axes are for both frames chosen as:
\begin{equation}
\hat{y}' = \left( p_{\eta}\times k_{\gamma}\right)/\left| p_{\eta}\times k_{\gamma}\right| 
\end{equation}
and the $x'$ axes lie in both systems in the reaction plane and are oriented such that
a right handed coordinate frame results. Angular distributions were constructed for the 
polar angle $\Theta_{\eta}$ of the $\eta$-meson in the photon - nucleon overall cm frame 
and for the polar ($\Theta_c$, $\Theta_h$) and azimuthal ($\Phi_c$, $\Phi_h$) angles of 
the pion in the canonical (`c') and helicity (`h') frames. Previous results for 
such angular distributions for photoproduction of $\pi^0\eta$ pairs off free protons 
are given in \cite{Kashevarov_09}.

\begin{figure}[thb]
\centerline{\resizebox{0.50\textwidth}{!}{%
  \includegraphics{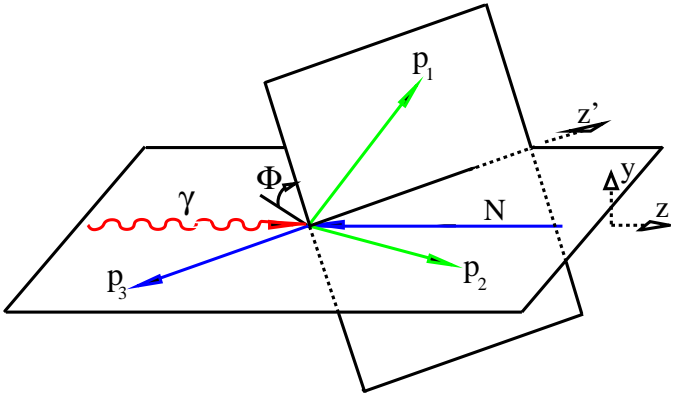}
}}
\caption{Vector and angle definitions in the cm system of incident photon ($\gamma$) and 
initial-state participant nucleon $N$. Particles $p_1$, $p_2$, and $p_3$ are some
permutation of the final-state participant nucleon $N'$, the pion, and the $\eta$ meson,
depending on the type of the asymmetry (see text). One plane is defined by the momentum
of the incident photon $\vec{k}$ and the momentum of particle $p_3$, the other by the
momenta of particles $p_1$ and $p_2$ (all momenta in the photon - nucleon cm system).
$\Phi$ is the angle between the planes.  
}
\label{fig:planes}       
\end{figure}
   
Final states with three particles (such as $\pi\pi N$ or $\pi\eta N$) can show an asymmetry, 
the beam-helicity asymmetry $I^{\odot}$, even when investigated with an unpolarized target 
and a circularly polarized photon beam. The definition of the asymmetry is shown in 
Fig.~\ref{fig:planes}.
Two planes are defined in the photon-nucleon cm system by the three final-state particles 
and the incident photon (or nucleon). The asymmetry due to the photon beam helicity
can be defined as a function of the angle $\Phi$ between the two planes by:
\begin{equation}
I^{\odot}(\Phi)=
	        \frac{d\sigma^{+}-d\sigma^{-}}{d\sigma^{+}+d\sigma^{-}}
	       =\frac{1}{P_{\gamma}}
                \frac{N^{+}-N^{-}}{N^{+}+N^{-}}\;,
\label{eq:circ}
\end{equation}
where $d\sigma^{\pm}$ are the differential cross sections for each of the two photon helicity 
states, $P_{\gamma}$ is the degree of circular polarization of the photons, and $N^{\pm}$ 
are the count rates for the two helicity states. The integration of the count rates over extended 
phase-space regions has to be corrected for detection efficiency effects. In the present 
analysis, results are presented for the choices $(p_1,p_2,p_3)$=$(\eta,\pi,N)$, $(\pi,N,\eta)$,
and $(\eta,N,\pi)$. The corresponding angles are labeled $\Phi_1$, $\Phi_2$, and $\Phi_3$
and the asymmetries are denoted as $I^{\odot}(\eta,\pi,N)$, $I^{\odot}(\pi,N,\eta)$, and
$I^{\odot}(\eta,N,\pi)$. Previous results for $I^{\odot}(\pi,N,\eta)$ for $\pi^0\eta$
pairs produced on free protons are given in \cite{Kashevarov_10}. 

Due to parity conservation, all asymmetries must obey the condition: 
\begin{equation}
I^{\odot}(\Phi)=-I^{\odot}(2\pi-\Phi),
\label{eq:sym1}
\end{equation} 
and can be expanded into sine series:
\begin{equation}
I^{\odot}(\Phi)=\sum_{n=1}^{\infty}A_{n}\mbox{sin}(n\Phi)
\label{eq:coeff}
\end{equation}
which can be fitted to the data. Test fits produced no $A_i, i\geq 3$ results significantly 
different from zero. This was due to the limited statistical 
precision. Therefore, the final fits were restricted to the $A_1$ and $A_2$ coefficients.

\section{Experimental setup}
\label{sec:setup}
The experimental setup and all relevant parameters have already been discussed in detail 
in several publications 
(see \cite{Oberle_13,Oberle_14,Dieterle_15,Kaeser_15,Werthmueller_13,Werthmueller_14,Dieterle_14}
for the measurements with deuterium targets and \cite{Pheron_12,Witthauer_13} for the $^3$He
data), which used the same data sets. Therefore, only a short summary is given here.
The experiments were carried out at the Mainz MAMI accelerator \cite{Herminghaus_83,Walcher_90}
using a quasi-monochromatic photon beam with energies up to $\approx$1.4 GeV from
the Glasgow tagged photon spectrometer \cite{Anthony_91,Hall_96,McGeorge_08}. The electrons were 
longitudinally polarized and this polarization is transferred in the brems\-strahlung process to 
circular polarization of the photons. The degree of circular polarization of the photon beam is
energy dependent and is related to the degree of linear polarization of the electron beam by the
transfer function given by Olsen and Maximon in \cite{Olsen_59}. The polarization degree of the 
electron beam was measured by Mott and M$\o$ller scattering (typical values were in the range
60 - 85\%, see \cite{Oberle_13,Oberle_14}).

In total, results from three different beam times with liquid deuterium targets, one
beam time with a liquid hydrogen target, and one beam time with a liquid $^3$He
target, were analyzed for the present results. Their main parameters are listed in 
Table \ref{tab:beam}. The target cells were Kapton cylinders (a Mylar cylinder for the
liquid $^3$He) of $\approx$ 4 cm diameter. Target lengths, densities, and the beam
parameters of the different beam times are summarized in Table~\ref{tab:beam}. 

\begin{table}[hhh]
\begin{center}
  \caption[Summary of data sets]{
    \label{tab:beam}
     Summary of data samples. Target type ($LD_2$: liquid deuterium, $LH_2$: 
     liquid hydrogen, $L^3He$: liquid $^3$He), target length 
     $d$ [cm], target surface density $\rho_s$ [nuclei/barn], 
     electron beam energy $E_e^-$ [MeV],
     degree of longitudinal polarization of the electron beam $P_{e^-}$ [\%].
}
\vspace*{0.3cm}
\begin{tabular}{|c|c|c|c|c|}
\hline
Target & d [cm] & ~$\rho_s$ [barn$^{-1}]~$ & 
$E_{e^-}$ [MeV] & $P_{e^-}$ [\%]\\
\hline\hline
  $LD_2$  & 4.72 & 0.231$\pm$0.005 & 1508 & 61$\pm$4\\
  $LD_2$  & 4.72 & 0.231$\pm$0.005 & 1508 & 84.5$\pm$6\\
  $LD_2$  & 3.00 & 0.147$\pm$0.003 & 1557 & 75.5$\pm$4\\
  $LH_2$  & 10.0 & 0.422$\pm$0.008 & 1557 & 75.5$\pm$4\\ 
  $L^3He$ & 5.08 & 0.073$\pm$0.005 & 1508 & - \\    
\hline
\end{tabular}
\end{center}
\end{table}

\begin{figure}[htb]
\centerline{\resizebox{0.49\textwidth}{!}{%
  \includegraphics{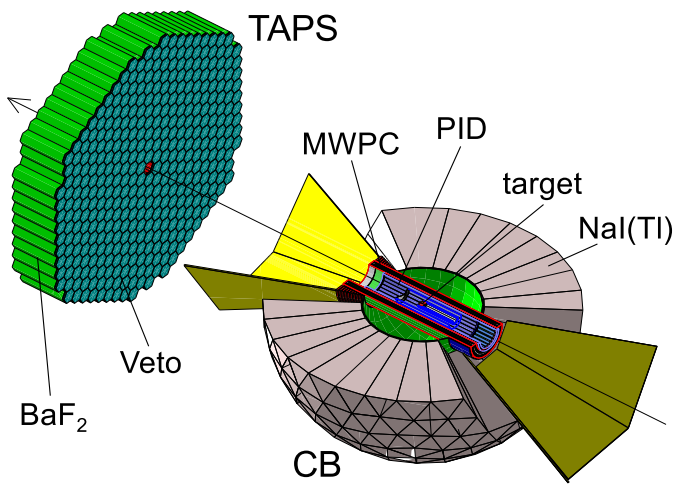}
}}
\caption{Setup of the electromagnetic calorimeter combining the
Crystal Ball and TAPS (left hand side) detectors. Only the lower half-shell
of the Crystal Ball detector is shown.
Detectors for charged particle identification are mounted in the Crystal Ball 
(PID and MWPC) and in front of the TAPS forward wall (TAPS Veto-detector).
}
\label{fig:calo}       
\end{figure}

Recoil nucleons, charged pions, and photons from the decay of the neutral mesons  
were detected in an almost $4\pi$ solid angle electromagnetic calorimeter, supplemented 
with detectors for charged particle identification (see Fig.~\ref{fig:calo}). 
The main components of this setup were the Crystal Ball (CB) detector \cite{Starostin_01} 
comprising 672 NaI crystals and a hexagonal forward wall constructed from 384 BaF$_2$ 
modules of the TAPS device \cite{Novotny_91,Gabler_94}. The CB covered polar angles from 
20$^{\circ}$ to 160$^{\circ}$ and the TAPS forward wall covered polar angles down to 
$\approx$5$^{\circ}$. All TAPS modules were equipped with individual plastic scintillators 
for charged particle identification in front of the crystals. The target was placed in the 
center of the CB and surrounded by a detector for charged particle identification (PID) 
\cite{Watts_04}. 

\section{Data analysis}
\label{sec:ana}
The data analysis for the final states summarized in this paper has already been discussed
in \cite{Kaeser_15}, where the total cross sections were presented. More details have 
been given for the almost identical analyses of the $N\pi^0\pi^0$ and $N\pi^0\pi^{\pm}$ final
states \cite{Oberle_14,Dieterle_15} from the same data set. Therefore, only a brief summary 
is given here.

\begin{figure*}[thb]
\centerline{\resizebox{0.97\textwidth}{!}{
  \epsfig{file=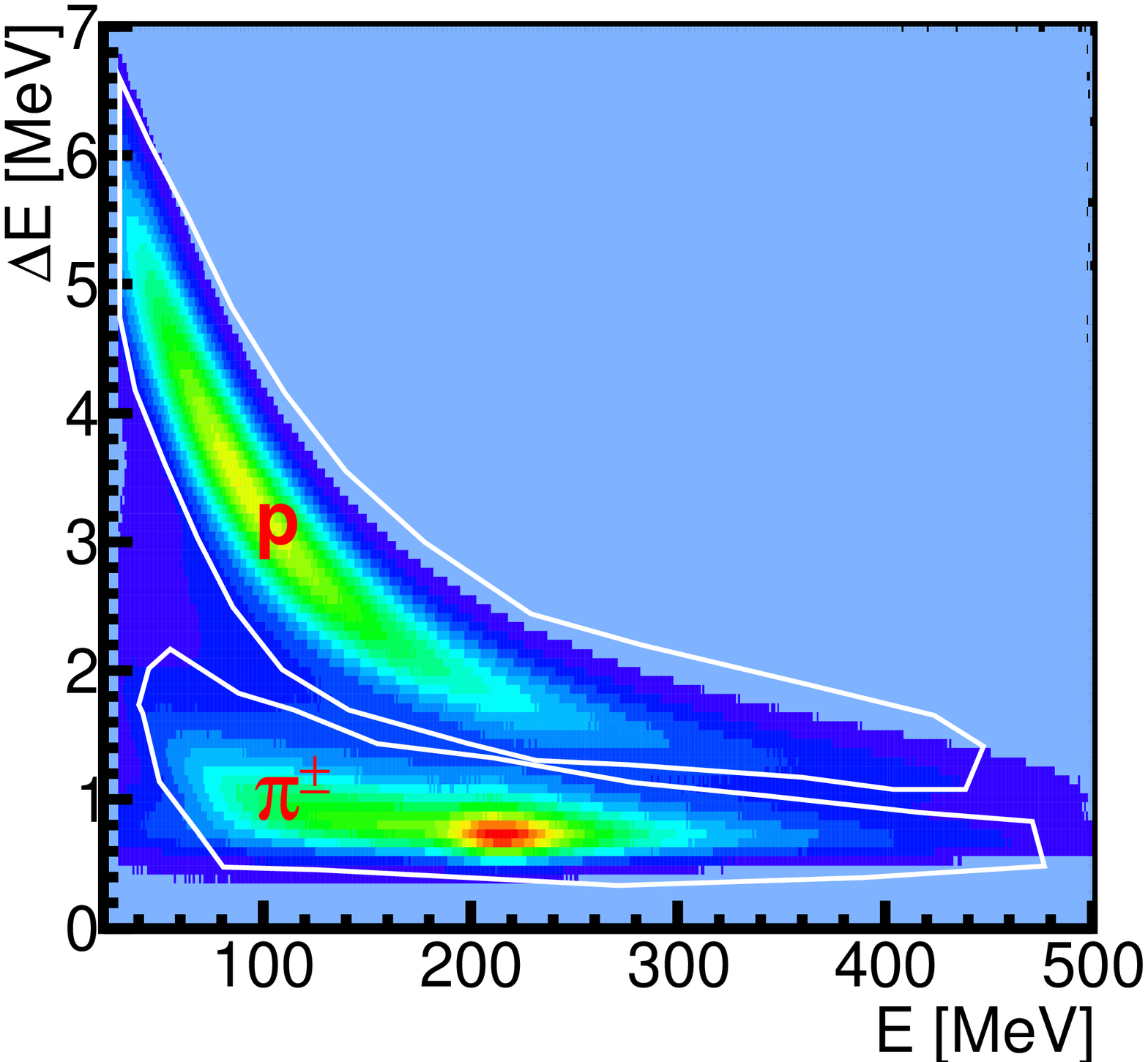,scale=0.33}
  \epsfig{file=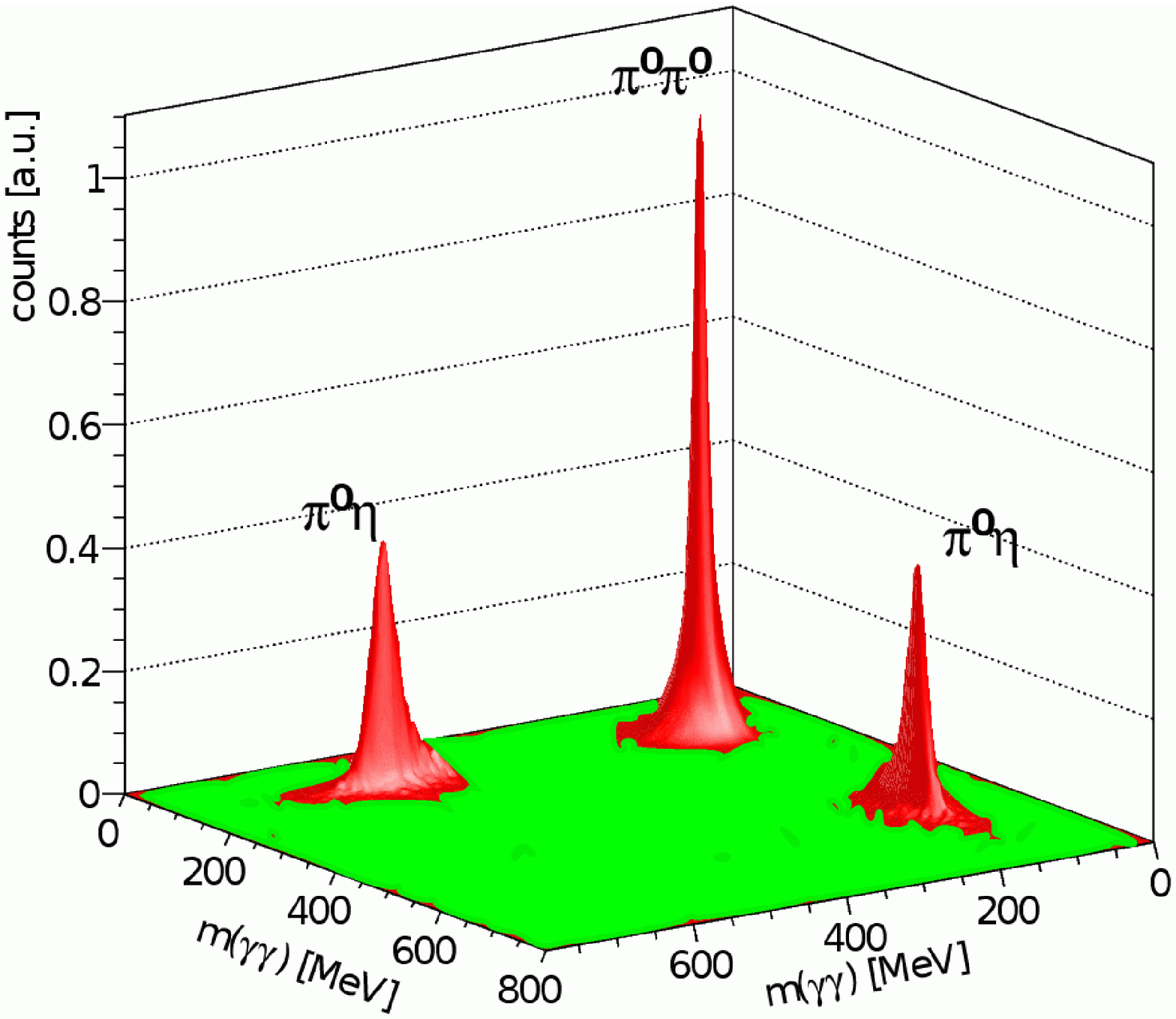,scale=0.33}
  \epsfig{file=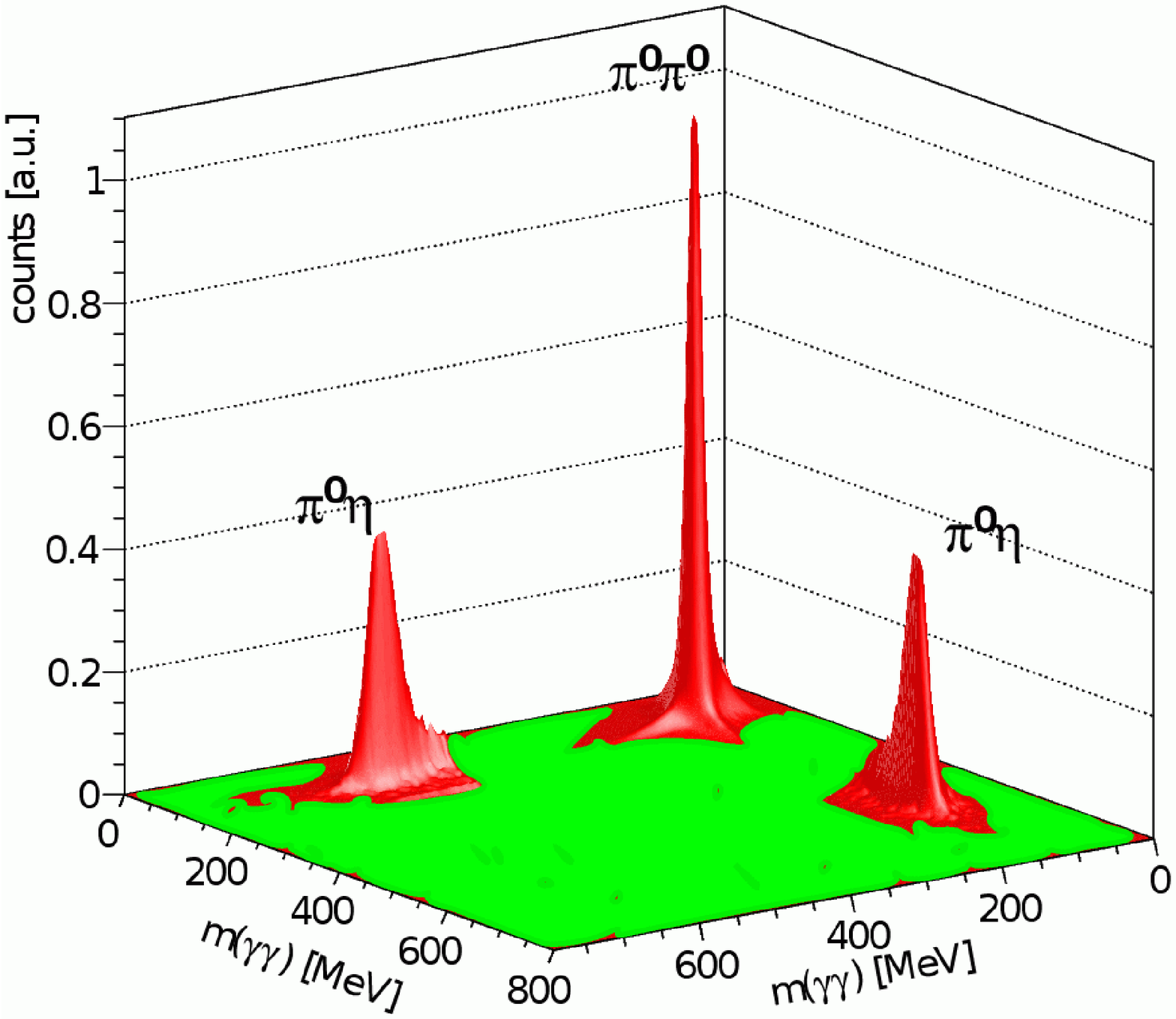,scale=0.33}  
}}
\caption{Left hand side \cite{Kaeser_15}: Identification of protons and charged pions in 
CB with a $\Delta E$ (energy depositon in PID) versus $E$ (total energy measured with CB)
analysis. White lines indicate the accepted areas. 
Center (right hand side): two-dimensional invariant-mass distributions for events 
with four photons in coincidence with recoil protons (recoil neutrons). 
The regions around the $\pi^0\eta$ peaks are scaled up by a factor of 50.} 
\label{fig:iden} 
\end{figure*}

In the first step of the analysis, the charged particle detectors in front of TAPS and inside 
the CB were used to classify the hits in the calorimeters as `charged' or `neutral'. 
Subsequently, a $E-\Delta E$ analysis, comparing the energy deposition in the PID to the full
energy of the particle measured with the CB was used to separate protons and
charged pions (see Fig.~\ref{fig:iden}, left hand side). Charged pions in TAPS were not 
analyzed because the contamination with the more abundant protons was substantial. 
This means that a small part of the reaction phase space (charged pions at polar angles 
smaller than 20$^{\circ}$) was excluded from the analysis (this was taken into account for 
the simulation of the detector acceptance and efficiency). Recoil protons and neutrons in TAPS 
were identified with a time-of-flight versus energy and a pulse-shape analysis. 

\begin{table}[h]
\begin{center}
\caption{Selected event classes for the cross sections $\sigma_p$ (coincident with 
recoil protons), $\sigma_n$ (coincident with recoil neutrons), and $\sigma_{\rm incl}$ 
(no condition for recoil nucleons) for $\pi\eta$-pairs with neutral and charged pions. 
$n$ and $c$ denote neutral and charged hits in the calorimeter (distinguished 
by the response of the charged-particle detectors).
} 
\label{tab:events}       
\begin{tabular}{|c|c|c|c|}
\hline\noalign{\smallskip}
& $\sigma_p$ & $\sigma_n$ & $\sigma_{\rm incl}$\\
\hline
$\pi^0\eta$     & 4$n$\&1$c$ & 5$n$  & 4$n$ or 5$n$ or (4$n$\&1$c$)\\
$\pi^{\pm}\eta$ & 2$n$\&2$c$ & 3$n$\&1$c$ & (2$n$\&1$c$) or (2$n$\&2$c$) or (3$n$\&1$c$)\\
\hline
\end{tabular}
\end{center}
\end{table}

The events characterized in Table~\ref{tab:events} were then accepted for the analysis 
of (quasi)-free production off protons ($\sigma_p$), off neutrons ($\sigma_n$),
and the inclusive reaction off the deuteron ($\sigma_{\rm incl}$), for which recoil nucleon 
detection was not required, but allowed. The inclusive cross section has only been used to 
check the relation:
\begin{equation}
\sigma_{\rm incl}\approx\sigma_p +\sigma_n+(\sigma_d), 
\label{eq:sum}
\end{equation}
where the (small) coherent cross section $\sigma_d$ measured in coincidence with recoil
deuterons contributes only for $\pi^0\eta$ pairs. As shown in \cite{Kaeser_15}, 
maximum deviations from Eq.~\ref{eq:sum} are below 5\%, which limits possible
uncertainties for the detection of recoil nucleons ($\sigma_p$, $\sigma_n$, and $\sigma_d$
depend on them, but $\sigma_{\rm incl}$ does not).

Photons and neutrons cannot be distinguished in the CB (the flight path is too short for
time-of-flight versus energy analysis, there is no pulse-shape analysis, and cluster-size
distributions of electromagnetic showers and energy depositions from neutrons in the CB
do not allow for an event-by-event separation (see \cite{Dieterle_15,Kaeser_15,Werthmueller_14})).
Therefore, neutral hits were assigned with a $\chi^2$ analysis to photons and neutrons.
For events with three or five neutral hits, the invariant masses of all possible pair 
combinations were compared to the mass of the $\pi^0$, or the $\pi^0$ and $\eta$ masses, 
respectively. For events with $n_m$ neutral mesons ($n_m$ = 1 for 
$\pi^{\pm}\eta$ final states, $n_m$ = 2 for $\pi^0\eta$) $\chi^2$ was defined by
\begin{equation}
\label{eq:chi2}
\chi^{2}(k) = \sum_{i=1}^{n_m}\left 
(\frac{m_{\pi^0,\eta}-m_{i,k}}{\Delta m_{i,k}}\right)^{2} ~~{\rm with}~~ k=1,..,n_p,
\end{equation} 
where the $m_{i,k}$ are the invariant masses of the $i$-th pair in the $k$-th permutation 
of the hits and $\Delta m_{i,k}$ is the corresponding uncertainty computed 
event-by-event from the experimental energy and angular resolution. 
In order to suppress combinatorial background, not only the hypotheses of $\pi^0\eta$ pairs
(for events with five neutrals) or $\eta$ mesons (for events with three neutrals)
but also those of $\pi^0\pi^0$ pairs, or single $\pi^0$ production were tested.
In all cases, only the combination with the minimum $\chi^2$ was selected for further 
analysis. A two-dimensional invariant mass spectrum of events with four photons is shown
in Fig.~\ref{fig:iden}, (right hand side), and details are discussed in \cite{Kaeser_15}.

\begin{figure}[htb]
\centerline{\resizebox{0.49\textwidth}{!}{%
  \includegraphics{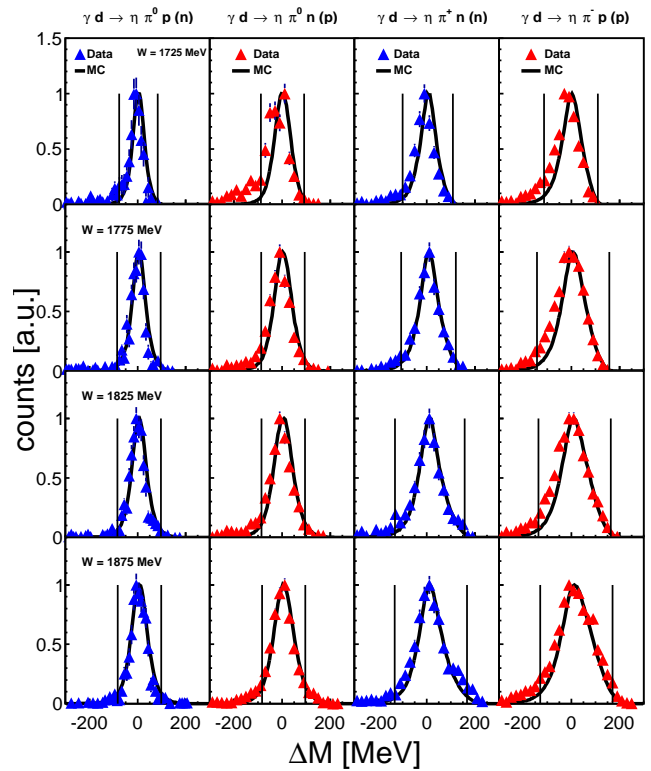}
}}
\caption{Missing mass distributions. From top to bottom: different values of $W$ 
corresponding to the ranges for which differential cross sections and asymmetries were
extracted (center values $\pm$25~MeV). Columns from left to right: reactions 
$\gamma d\rightarrow\eta\pi^0 p(n)$,
$\gamma d\rightarrow\eta\pi^0 n(p)$, $\gamma d\rightarrow\eta\pi^+ n(n)$, and
$\gamma d\rightarrow\eta\pi^- p(p)$ (in parenthesis undetected spectator nucleon).
Triangles are data, and solid curves are MC simulation of the signals. 
Vertical lines: ranges of accepted events.
}
\label{fig:mima}       
\end{figure}

In the final step of the reaction identification, residual background was removed by an analysis
of the coplanarity of the final state particles and the missing mass of the reaction when
the recoil nucleon (although detected) was treated as a missing particle (see
\cite{Kaeser_15} for details). The coplanarity analysis is based on the fact that in the 
cm system (apart from Fermi smearing) the momentum vector of the combined $\pi\eta$ 
system must be back-to-back with the recoil nucleon momentum, so that the azimuthal angles 
of the nucleon and the combined  $\eta\pi$ laboratory momentum must differ by 
180$^{\circ}$. This cut suppresses also the rare events were due to large Fermi momenta the 
spectator nucleon is detected instead of the participant nucleon. The missing mass analysis uses the 
incident photon energy and the four momenta of the two mesons to kinematically reconstruct 
the mass of the `missing' particle (the detected participant recoil nucleon, Fermi motion is 
neglected) and compares it to the mass of the nucleon via: 
\begin{equation}
\label{eq:2pimiss}
\Delta M = \left|P_{\gamma}+P_{N}-P_{\pi}-P_{\eta}\right| -m_N\ ,
\end{equation}
where $m_N$ is the nucleon mass, $P_{\gamma}$ is the four-momentum of the incident photon, 
$P_N$ is the four momentum of the initial state nucleon (assumed at rest), and $P_{\pi}$ and
$P_{\eta}$ are the four momenta of the two mesons. Typical spectra for all four reactions 
for the energy ranges of interest are summarized in Fig.~\ref{fig:mima}. They are almost 
background free and the shape of the distributions is in good agreement with Monte Carlo
simulations. The tails of the distributions were rejected in order to remove small residual 
backgrounds and kinematically poorly reconstructed events. 

Absolute cross sections were extracted from the measured yields, as described in  
\cite{Kaeser_15} (more details are given in \cite{Dieterle_15,Werthmueller_14}
for other reactions), from the target surface
densities, the photon flux, the meson decay branching ratios, and the simulated
(Geant4 \cite{GEANT4}) detection efficiency of the experimental setup. For the asymmetries, 
the polarization degree of the photon beam also matters, which was discussed in 
\cite{Oberle_13,Oberle_14} in context of the photoproduction of pion pairs.  

For the total cross sections, which have been summarized in \cite{Kaeser_15}, two
different analyses were done. Excitation functions were obtained as a function of the
photon energy measured with the tagging spectrometer. This analysis suffered from Fermi smearing.
The other analysis reconstructed the final state total energy $W$ from the incident photon energy, 
the four momenta of the mesons, and the polar and azimuthal angles of the detected recoil nucleon. 
This analysis eliminated the effects from Fermi motion, but introduced effects from angular and 
energy resolution of the calorimeter. For the differential cross sections and asymmetries, 
only the second method was used because Fermi smearing obscures these observables 
too much. Note that the kinematic reconstruction of the final state is only exact for
the deuterium target. For the $^3$He target, it is based on the (fairly good) approximation that 
the two spectator nucleons have no relative momentum (for details see \cite{Witthauer_13}).   

Systematic uncertainties have been discussed in detail in
\cite{Oberle_13,Oberle_14,Dieterle_15,Werthmueller_14}, which analyzed the same data for
other reaction channels, and in \cite{Kaeser_15} specifically for the production of 
$\pi\eta$ pairs. The total overall normalization uncertainty (photon flux, target density) 
was estimated to be between 5\% (quadratic addition) and 7\% (linear addition).
Uncertainties from analysis cuts including simulation of the detection efficiency, 
but excluding the recoil nucleon detection, were in the range of 5 - 10\% (larger uncertainties
for charged pions). The detection efficiency has been simulated with different event generators 
using reaction phase space and the 
$\Delta^{\star}\rightarrow \eta\Delta\mbox{(1232)}\rightarrow \eta\pi^0 p$ decay chain.
The differences were small and the results from the sequential decay, which is in excellent
agreement with the measured invariant mass distributions, were used. In addition to Monte Carlo
simulations, the recoil nucleon detection efficiencies have been also experimentally investigated
with reactions such as $\gamma p\rightarrow p\eta$, $\gamma p\rightarrow p\pi^0$, 
$\gamma p\rightarrow p\pi^0\pi^0$ for recoil protons and $\gamma p\rightarrow n\pi^+\pi^0$
for recoil neutrons \cite{Dieterle_15,Werthmueller_14}. The systematic uncertainty of these
analyses has been estimated in \cite{Werthmueller_14} at the 10\% level, but the good agreement 
between the sum of the exclusive cross sections and the inclusive results for many different reaction
channels suggest that this is an upper limit. Most of the uncertainties cancel
in the isospin ratios discussed in \cite{Kaeser_15}. The systematic uncertainty of the 
asymmetries is dominated by the precision of the measurement of the polarization degree of the
electron beam (see Table~\ref{tab:beam}).

\section{Results}
\label{sec:results}
\begin{figure*}[thb]
\centerline{\resizebox{0.79\textwidth}{!}{%
  \includegraphics{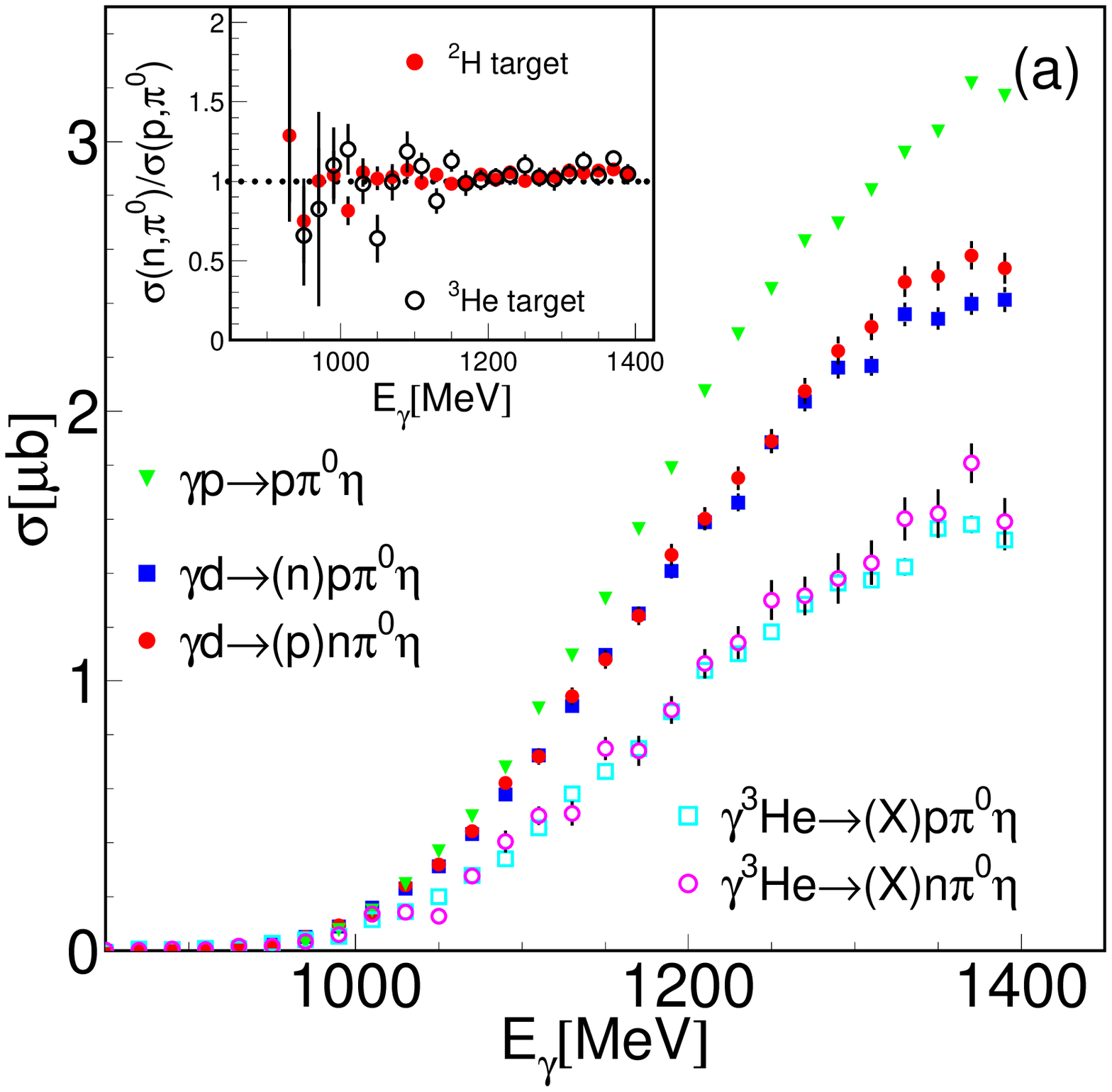}
  \hspace*{1cm}
  \includegraphics{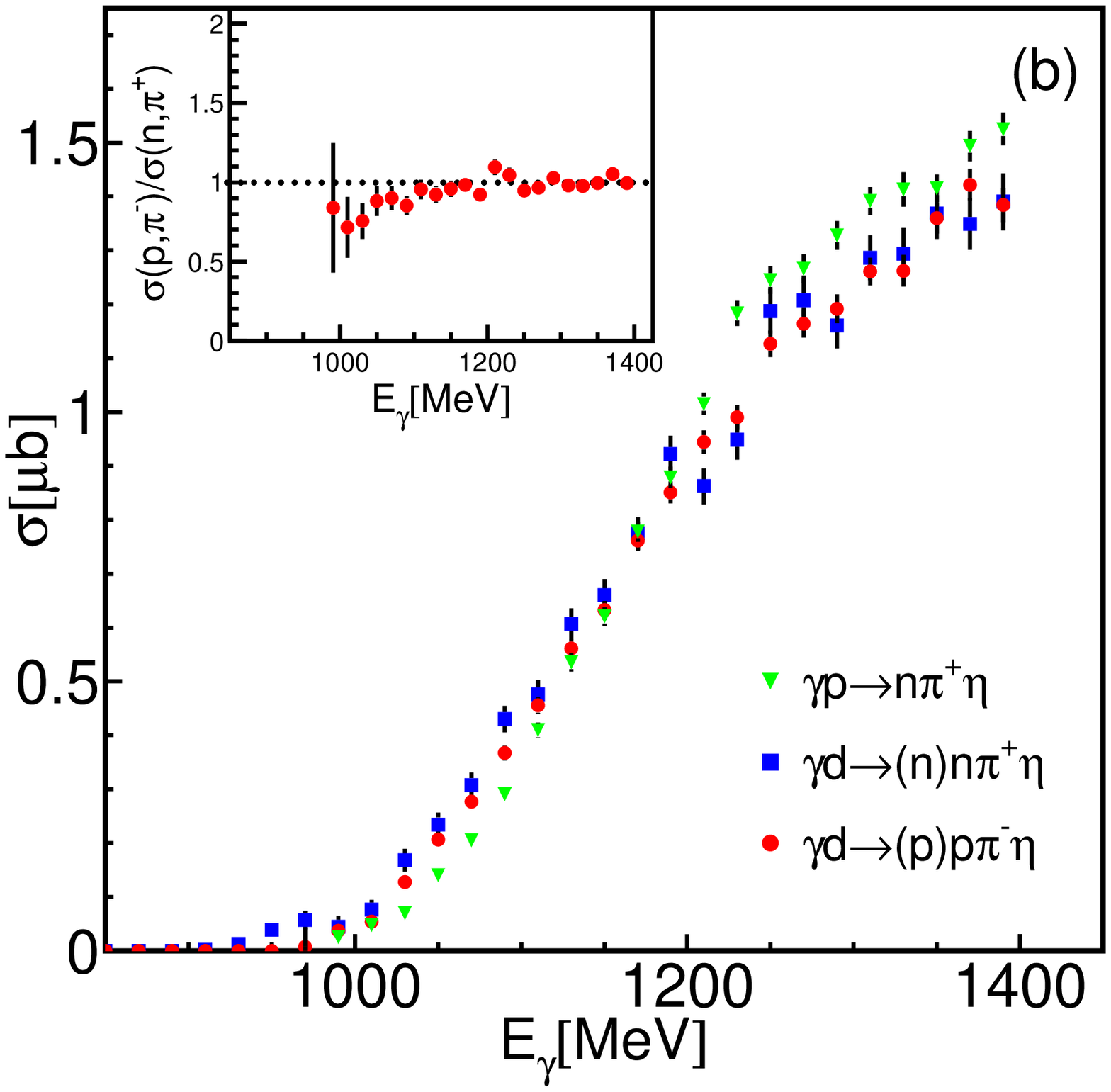}  
}}
\centerline{\resizebox{0.79\textwidth}{!}{%
  \includegraphics{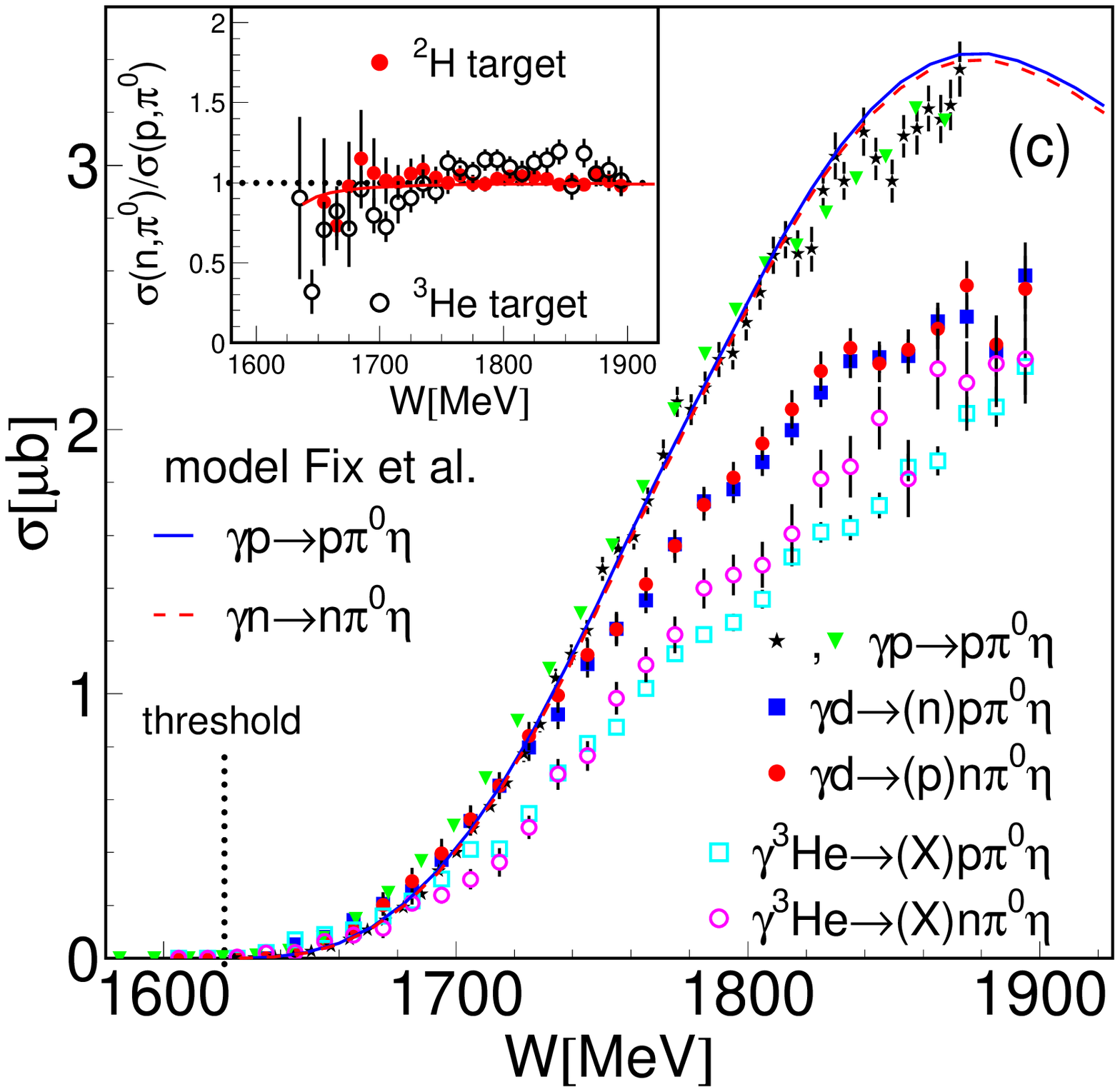}
  \hspace*{1cm}
  \includegraphics{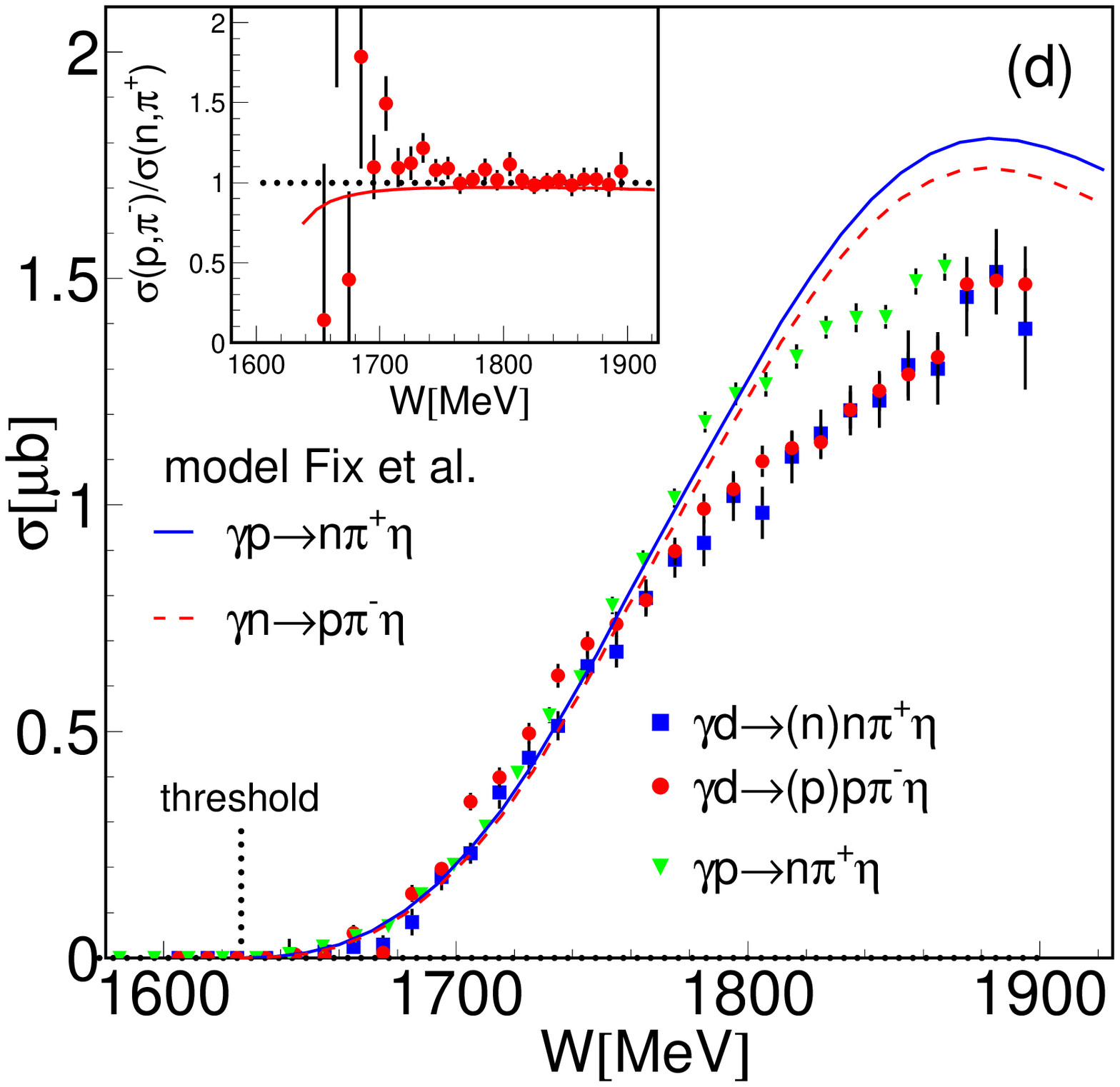} 
}} 
\vspace{0.3cm} 
\centerline{\resizebox{0.79\textwidth}{!}{%
  \includegraphics{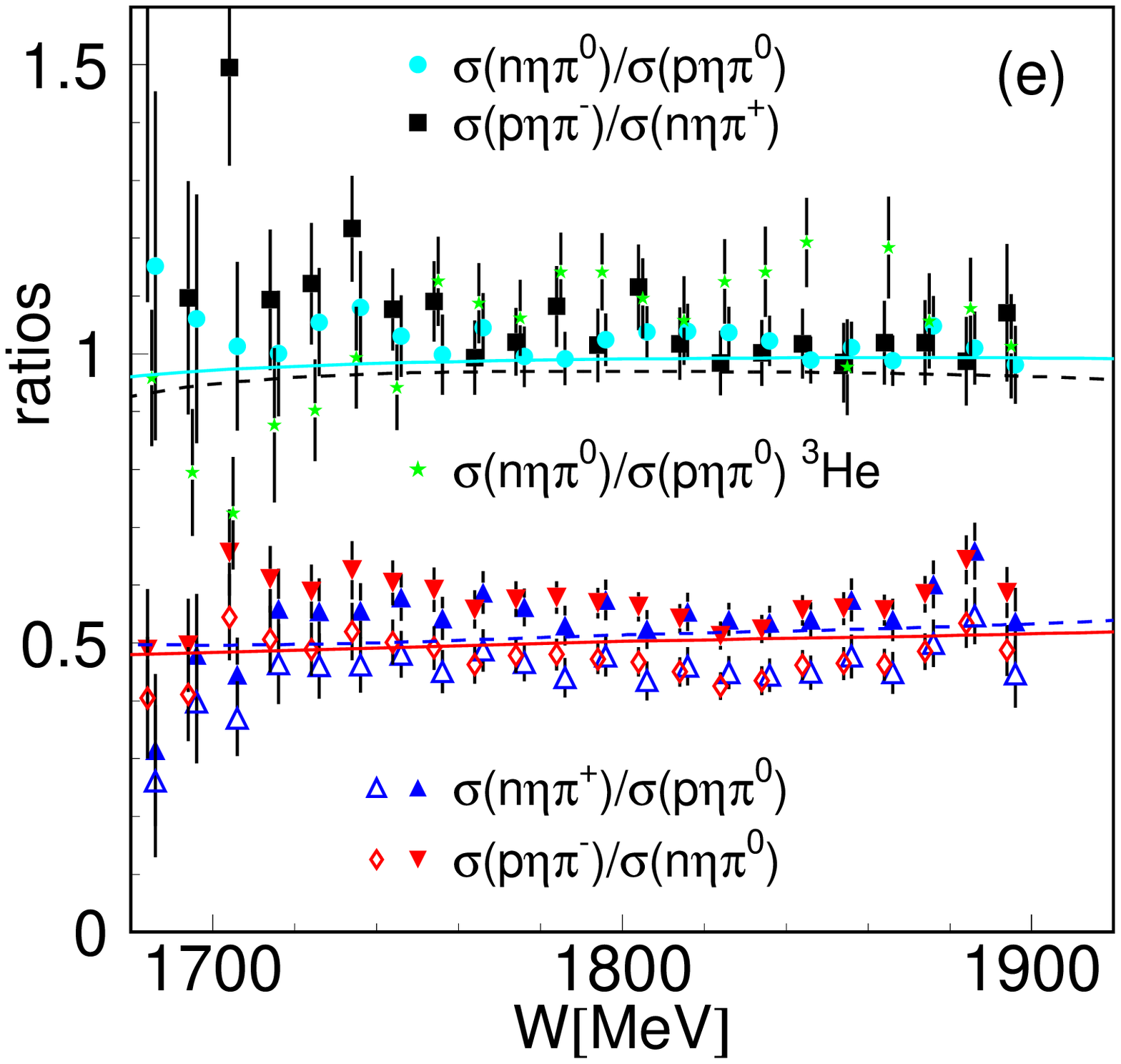}
  \hspace*{1cm} 
  \includegraphics{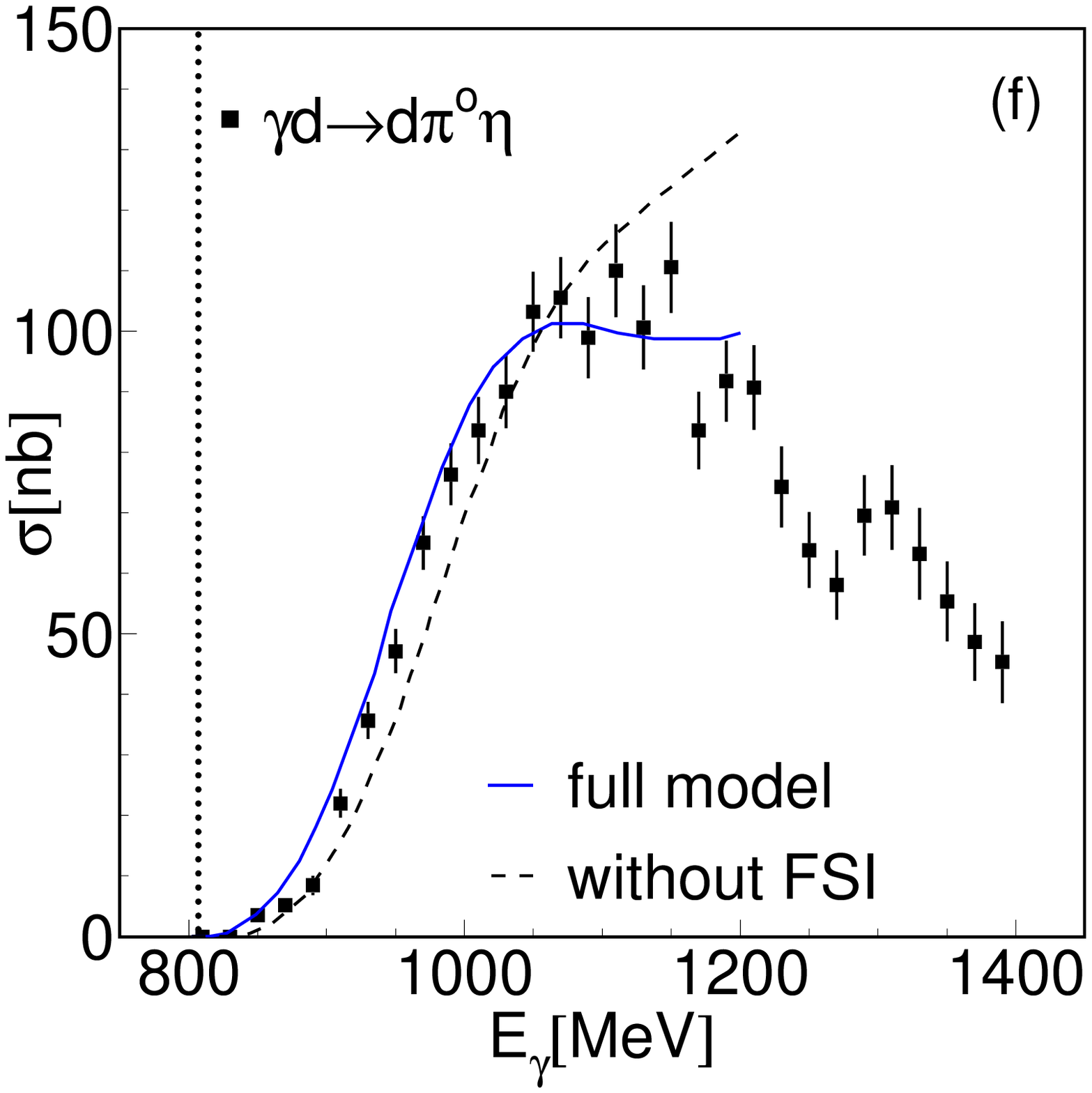}  
}}
\caption{Total cross sections for $\gamma N\rightarrow N\pi\eta$ reactions. (a) and (b) show
the results as a function of photon energy, (c) and (d) as a function  of final state 
invariant mass. Figures on the left hand side ((a) and (c)) correspond to the $\pi^0\eta$ final state, 
figures on the right hand side ((b) and (d)) to the $\pi^{\pm}\eta$ final state. 
The different reactions are indicated in the figures. 
Previous results for $\gamma p\rightarrow p\pi^0\eta$ in (c), black stars, are from \cite{Kashevarov_09}. 
The inserts show the ratio of quasi-free production off neutrons and protons from deuterium targets 
and for the $\pi^0\eta$ final state also for a $^3$He target. 
(e) summarizes the results for the ratios of the isospin channels. For the ratios involving
charged and neutral pions the solid symbols are as measured and the open symbols are re-scaled
for FSI effects (see text). The curves represent the model results from \cite{Fix_10}. 
(f) shows the data for the $\gamma d\rightarrow d\pi^0\eta$ reaction \cite{Kaeser_15}
with model results from \cite{Egorov_13}. Only the statistical uncertainties are shown. 
}
\label{fig:tpi0}       
\end{figure*}

Total cross sections and their ratios for the different isospin channels have already been 
presented in \cite{Kaeser_15}. Here, the main results are summarized. All differential
distributions have been normalized by the absolute scale of the total cross sections
for easier comparison of the shapes. 
Fig.~\ref{fig:tpi0} shows the total cross sections for the four final states as a function
of the incident photon energy (Figs.~\ref{fig:tpi0}(a),(b)) and as a function of
the final state invariant mass $W$ (Figs.~\ref{fig:tpi0}(c),(d)). Only the exclusive quasi-free
cross sections measured in coincidence with recoil nucleons are shown and compared to previous 
and present results for a free proton target. A comparison of the inclusive cross sections to the
sum of the exclusive ones was made in Ref.~\cite{Kaeser_15} to demonstrate the validity of 
Eq.~\ref{eq:sum}.

The total cross section for the $\gamma p\rightarrow p\pi^0\eta$ reaction is in very good 
agreement with the previous measurement from Ref.~\cite{Kashevarov_09} (note that the results
from CBELSA \cite{Gutz_14} were originally $\approx$15\% lower, but have been renormalized
in \cite{Gutz_14} to \cite{Kashevarov_09} because of their larger systematic uncertainty). 
For the $\pi^0\eta$ final state, 
significant effects from final state interactions (FSI) have been observed. When analyzed as 
a function of incident photon energy, the quasi-free cross section for production off protons 
is roughly 75\% for protons bound in the deuteron and only 50\% for protons bound in $^3$He, 
both relative to the cross section for the free proton target. 
Analyzed as a function of final state $W$, the quasi-free cross sections are 75\% and 60\% 
relative to the free-proton cross section. For the results analyzed as a function of $E_{\gamma}$, 
deviations between free and quasi-free reactions are due to FSI effects {\it and} Fermi smearing.
Fermi motion effects have been removed from the $W$ data (some residual effects from the
reconstruction may be present in the imediate vicinity of the production threshold, which is,
however, not discussed in detail). Since the Fermi motion effects are larger for the $^3$He nucleus, 
the observed behavior is plausible. 

The FSI effects are smaller for the $\pi^{\pm}\eta$ final states (cross section for protons bound 
in the deuteron reduced to $\approx$90\%; it was not analyzed for the $^3$He target). This behavior 
is similar to the photoproduction of single pions. Also in this case FSI effects are larger for 
neutral pions due to the difference between nucleon-nucleon interactions in the $np$ system (which 
can be bound) compared to the $nn$ and $pp$ systems \cite{Krusche_03}. However, a quantitative 
understanding of the FSI effects is not yet available. The comparison of differential spectra discussed
below seems to indicate that the FSI effects manifest themselves mainly in the absolute scale of 
the cross sections. 

The ratios of the different isospin channels agree with the expectation for the 
$\Delta^{\star}\rightarrow \eta\Delta\mbox{(1232)}\rightarrow \eta\pi^0 p$ decay chain 
given in Eq.~\ref{eq:isorel}. They are summarized in Fig.~\ref{fig:tpi0}(e).
These ratios could be influenced by FSI effects when FSI is different for recoil protons and recoil 
nucleons and/or for charged and neutral pions. For the $\pi^0\eta$ final state, the $\sigma_n/\sigma_p$
ratio (quasi-free nucleons) extracted from the deuterium and $^3$He targets are comparable, although
the absolute effect from FSI is much stronger for the helium target. This is an indication that
FSI for quasi-free neutrons and protons are similar and thus the quasi-free
neutron/proton ratios are a good approximation for the reactions with free nucleons. The situation
is different for the comparison of reactions with the same type of target nucleon, but different
charge states of the pions since, as discussed above, FSI effects are more significant for production 
of neutral pions than for charged pions. Fig.~\ref{fig:tpi0}(e) shows the ratios of
neutral/charged pion cross sections as measured (solid symbols) and rescaled by the ratio of
FSI effects observed for protons (75\%$\div$ 90\% = 0.83) (open symbols).
Within uncertainties, the results are in agreement with Eq.~\ref{eq:isorel}.

The results with the effects from Fermi motion corrected are compared in Figs.~\ref{fig:tpi0}(c),(d) 
to the predictions from \cite{Fix_10}. All results from this reference correspond to their fit solution (I), 
which is in best agreement with data for the $\gamma p\rightarrow p\pi^0\eta$ reaction (in particular 
for the beam asymmetry $\Sigma$). This is their basis solution with the strong $D_{33}$ dominance.
Their solution (II) has at high incident photon energies a larger contribution from the $\Delta(1920)3/2^+$
state and their solution (III) has an admixture of the $\Delta(1700)3/2^-\rightarrow \Delta(1232)\eta$ decay
in $d$-wave, which is neglected in solution (I) which assumes only $s$-wave decay of this state.  
The model has not been fitted to the present data, i.e. not 
to any data with neutrons in the initial state or charged pions in the final state. The model predictions 
are only valid for free proton and free neutron targets and do not include FSI effects. Therefore, all
differential spectra have been renormalized to the total cross sections so that the major FSI effects 
were eliminated in the comparison of measured data and predictions. 

\clearpage

\begin{figure}[thb]
\centerline{\resizebox{0.48\textwidth}{!}{%
  \includegraphics{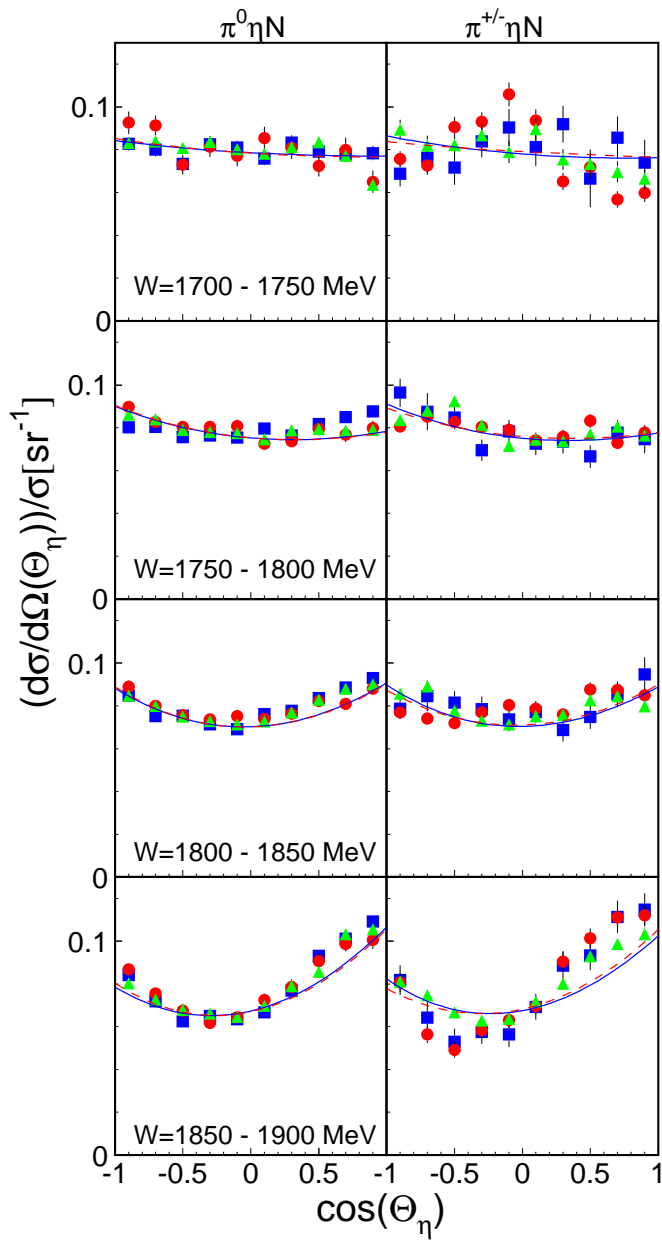}
}}
\caption{Distributions of the polar angles of the $\eta$ mesons in the photon - nucleon cm frame.
Left hand side: neutral pions, right hand side: charged pions. Ranges for invariant mass $W$ are 
given in the figure.
Initial states: (green) triangles: free protons, (blue) squares quasi-free protons and 
(red) circles: quasi-free neutrons. Curves for model results from \cite{Fix_10}, solid (blue) 
for target protons, dashed (red) target neutrons (proton and neutron predictions are almost 
identical).
}
\label{fig:theta_eta}       
\end{figure}

The model predicts similar results for protons and neutrons for the total magnitude of
the cross sections. The differences for charged pions are slightly larger than for neutral pions
but still below the 4\% level for all values of $W$. The predictions for the shape of the
differential distributions for protons and neutrons are even closer to each other,
as discussed below.

This is inherent to the model assumptions because in the initial electromagnetic photon-excitation 
process $\gamma N\rightarrow R$, only $I=3/2$ $\Delta$ resonances are taken into account for $R$ 
(with subsequent hadronic decay to $\Delta\eta$ or $N^{\star}\pi$ intermediate states). 
Isospin $I=1/2$ components contribute only via non-resonant background terms (Born-diagrams). 
Since the $\gamma N\Delta$ couplings are equal for protons and neutrons, the predicted proton and neutron
cross sections from resonance excitations are identical. 

The effects of the Born-terms are minor such that for the renormalized differential spectra in most figures, 
model results for proton and neutron targets are indistinguishable. Apart from small differences for the 
contributions of the Born-terms to reactions with neutral and charged pions, the relation between cross 
sections for different pion types is simply given by the Clebsch-Gordon coefficients. For the ratios of 
the total cross sections (for which the dominant FSI effects cancel), these predictions are clearly 
supported by the experimental results (see Figs.~\ref{fig:tpi0}~(c)-(e)).       

Also analyzed \cite{Kaeser_15} was the coherent production of $\eta\pi^0$ pairs off the deuteron:
\begin{equation}
\label{eq:coh}
\gamma+ d\rightarrow d +\eta +\pi^0\;.
\end{equation}
Since the deuteron has isospin $I=0$, the amplitude of this reaction is proportional 
to the sum of the amplitudes on protons and neutrons:
\begin{equation}
\label{eq:cohamp}
A(\gamma d\rightarrow d\pi^0\eta)\propto
A(\gamma p\rightarrow p\pi^0\eta)+A(\gamma n\rightarrow n\pi^0\eta).
\end{equation}
Inserting Eq.~(\ref{eq:iso}) into Eq.~(\ref{eq:cohamp}), one sees that in the coherent process 
of Eq.~(\ref{eq:coh}) the isoscalar excitation of $N^{\star}$ resonances is forbidden, 
so that only the isovector part of the $\gamma N\rightarrow N^{\star}$ transition 
and $I=3/2$ resonances can contribute. Thus the coherent reaction works 
as an isospin filter. In Fig.~\ref{fig:tpi0}(f), the data are compared to the model 
predictions from Ref.~\cite{Egorov_13}, where the elementary production operator from
\cite{Fix_10} (Solution(I)) was used. As mentioned above, the model \cite{Egorov_13} 
contains only the $\Delta$-like $I=3/2$ resonances as initial states, and the $I=1/2$ amplitude, 
coming only from the Born terms, is insignificant. Therefore, the good agreement between data and 
the predictions in Fig.~\ref{fig:tpi0}(f) is further evidence that 
the isospin decomposition is understood and that excitation of $N^{\star}$ resonances 
in the initial state of this reaction should not be significant.

It should be mentioned that contrary to single $\eta$ production, coherent production of $\eta\pi$
pairs is allowed for spin/isospin zero nuclei such as $^4$He. This reaction could thus be
used \cite{Krusche_15} for the search for $\eta$-mesic $^4$He similar to the use of $\eta$ production 
for $\eta$-mesic $^3$He \cite{Pheron_12,Pfeiffer_04}. 

For the quasi-free $\gamma N\rightarrow N\pi^0\eta$ and $\gamma N\rightarrow N'\pi^{\pm}\eta$ reactions, 
angular distributions, invariant mass distributions, and the helicity asymmetries $I^{\odot}$
have been analyzed. All differential spectra are shown for the $W$ ranges 1700 - 1750 MeV, 1750 - 1800 MeV,
1800 - 1850 MeV, and 1850 - 1900 MeV.

The distributions of the cm polar angles of the $\eta$-meson are summarized in Fig.~\ref{fig:theta_eta}.
For these spectra (and also for all following differential spectra), one should note that in general, the
statistical quality of the data is not as good with charged pions in the final state as for neutral
pions. 
\begin{figure*}[thb]
\centerline{\resizebox{0.96\textwidth}{!}{%
  \includegraphics{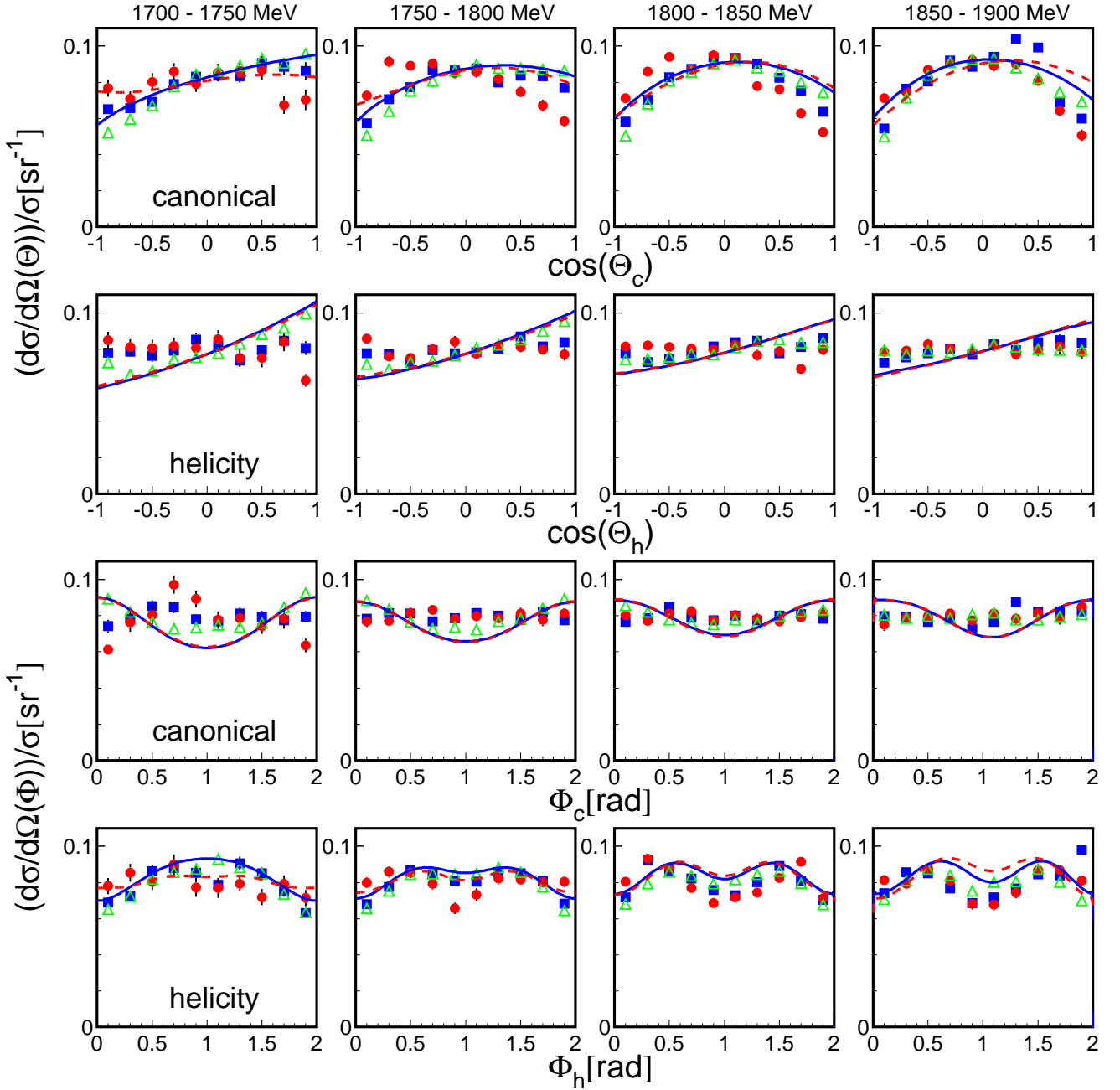}
}}
\caption{Angular distributions of the $\pi^0$ mesons from the $\gamma N\rightarrow N\eta\pi^0$ reactions
in the frames defined in Fig.~\ref{fig:angles} for different energy bins (given at top of the figure). 
$\Theta_c$, $\Phi_c$ ($\Theta_h$, $\Phi_h$) are the polar and azimuthal angles in the canonical and
helicity frames (Fig.~\ref{fig:angles}), respectively.
Notation for the experimental data same as in Fig.~\ref{fig:theta_eta}. 
Curves for model results from \cite{Fix_10}, solid (blue) for target protons, dashed (red) target neutrons 
(predictions for protons and neutrons almost identical for most figures). Only the statistical 
uncertainties are shown.
}
\label{fig:ang_pi0}       
\end{figure*}
This is because the total cross sections for charged pions are smaller by a factor of two and 
the detection efficiency for charged pions is also lower (in particular for low-energy pions). 
Furthermore, due to the larger detection efficiency for protons compared to neutrons (roughly a factor of 
three), the statistical precision for reactions involving recoil protons is better 
than for recoil neutrons. Thus for reactions with neutral mesons, more precise data were obtained for 
target protons (free or quasi-free) than for quasi-free neutrons, for charged pions it is the opposite. 

\begin{figure*}[htb]
\centerline{\resizebox{0.96\textwidth}{!}{%
  \includegraphics{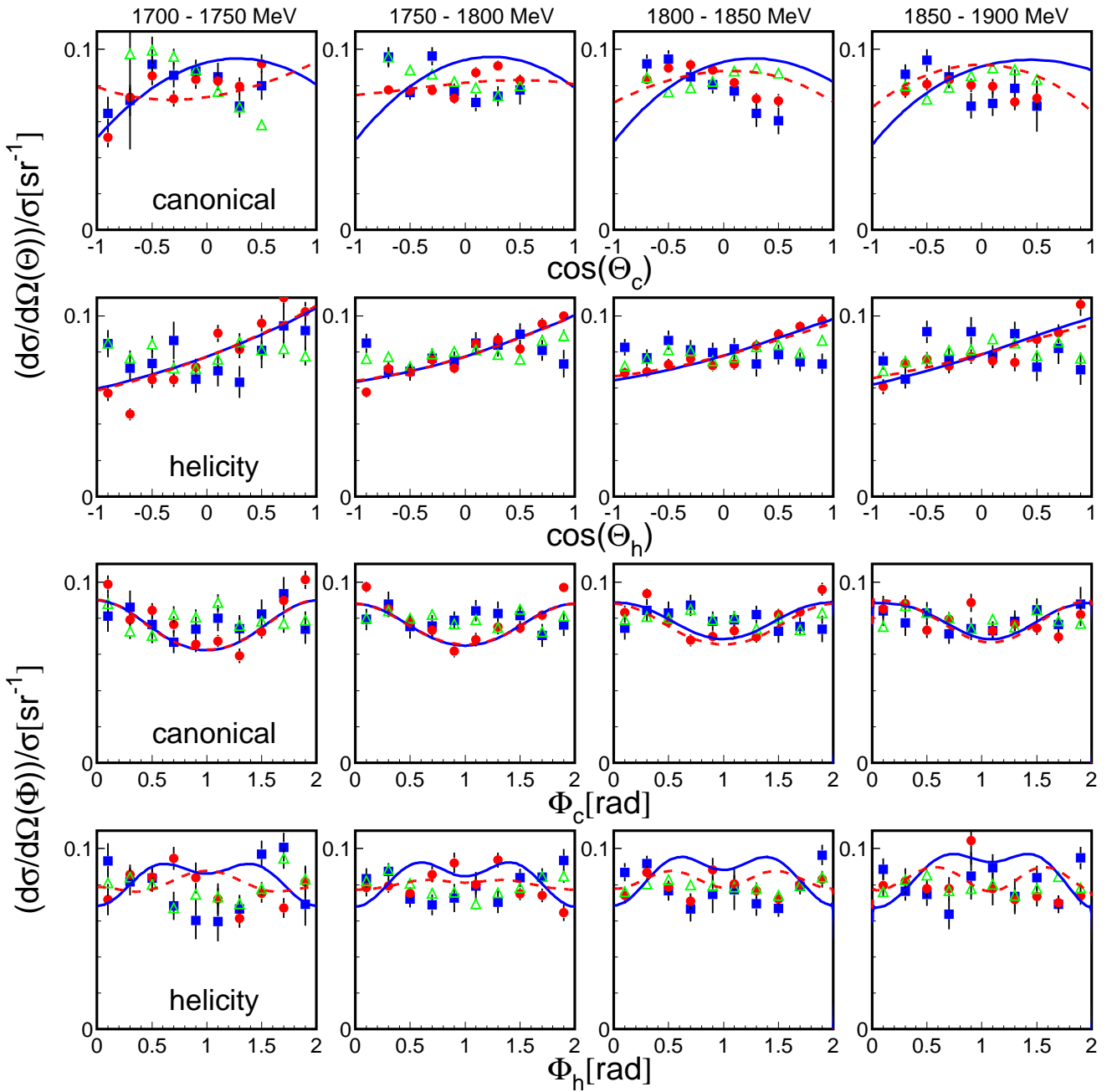}
}}
\caption{Same as in Fig.~\ref{fig:ang_pi0} for $\pi^{\pm}$ mesons from the 
$\gamma N\rightarrow N\eta\pi^{\pm}$ reactions.
}
\label{fig:ang_pic}       
\end{figure*}

The main observations for the $\eta$ polar angle distributions are: after renormalization by the 
total cross section, there are almost no differences between free and quasi-free proton data. This 
observation is true for almost all differential spectra and means that the FSI affects mainly the 
absolute scale of the cross section but has only a small effect on the shape of the spectra. As expected, 
the results for target neutrons and protons are similar because of the dominant excitation of 
$\Delta^{\star}$-resonances as doorway states. For the same reason, reactions with neutral and charged 
pions in the final state are also identical within uncertainties. 

The distributions are in good agreement in all iso\-spin channels (three of which have been measured 
for the first time) with the model predictions from \cite{Fix_10} assuming dominance of the 
$\Delta(1700)3/2^-\rightarrow\Delta(1232)3/2^+$ decay in the threshold region and, at higher  
photon energies, of the $\Delta(1940)3/2^-\rightarrow\Delta(1232)3/2^+$ decay. 
Some shape deviations between predictions and measured data appear only for charged pions at lowest and
highest incident photon energies. The predicted distributions are almost isotropic close to threshold 
where the $s$-wave decay of the $\Delta(1700)3/2^-$ to the $\Delta(1232)\eta$ intermediate state dominates. 
At higher photon energies the decay of this state to $\Delta(1232)\eta$ in relative $d$-wave and 
contributions from further initial $\Delta$ states may also become significant, in particular in 
interference terms. In solution (I) from \cite{Fix_10}, in addition to $\Delta(1700)3/2^-$, the $\Delta(1940)3/2^-$ 
state makes an important contribution. Further small contributions arise from the 
$\Delta(1905)5/2^+$ and $\Delta(1920)3/2^+$ resonances, while the $d$-wave decay of the $\Delta(1700)3/2^-$ 
and nucleon Born terms play almost no role. 

\begin{figure*}[thb]
\centerline{\resizebox{0.91\textwidth}{!}{%
  \includegraphics{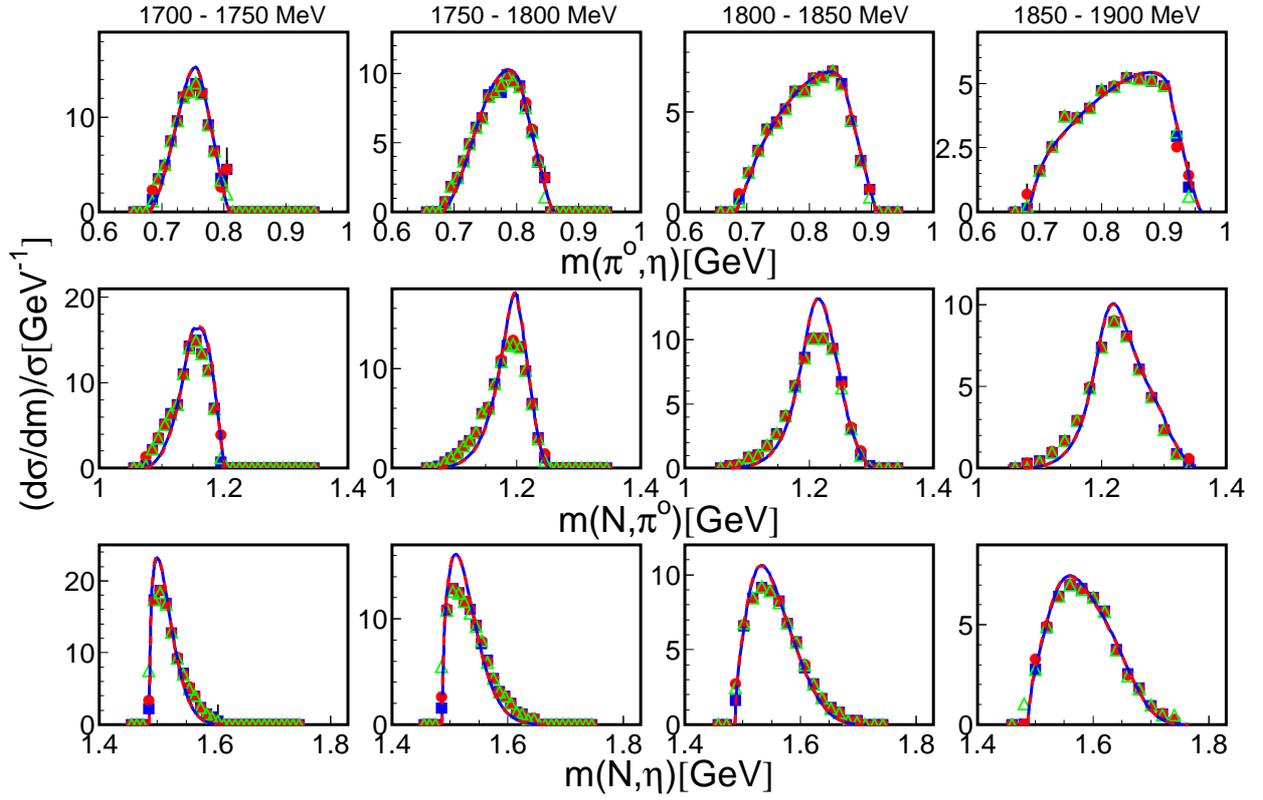}
}}
\caption{Invariant mass distributions for the $\pi^0\eta N$ final state. 
Notation is the same as in Fig.~\ref{fig:theta_eta}. 
Model predictions from \cite{Fix_10} are almost identical for target protons (blue, solid) 
and neutrons (dashed, red). Only the statistical uncertainties are shown.
Upper row: $\eta$ - pion invariant masses, central row: nucleon - pion, 
bottom row: nucleon - $\eta$.}
\label{fig:mpi0p}       
\end{figure*}

\begin{figure*}[htb]
\centerline{\resizebox{0.91\textwidth}{!}{%
  \includegraphics{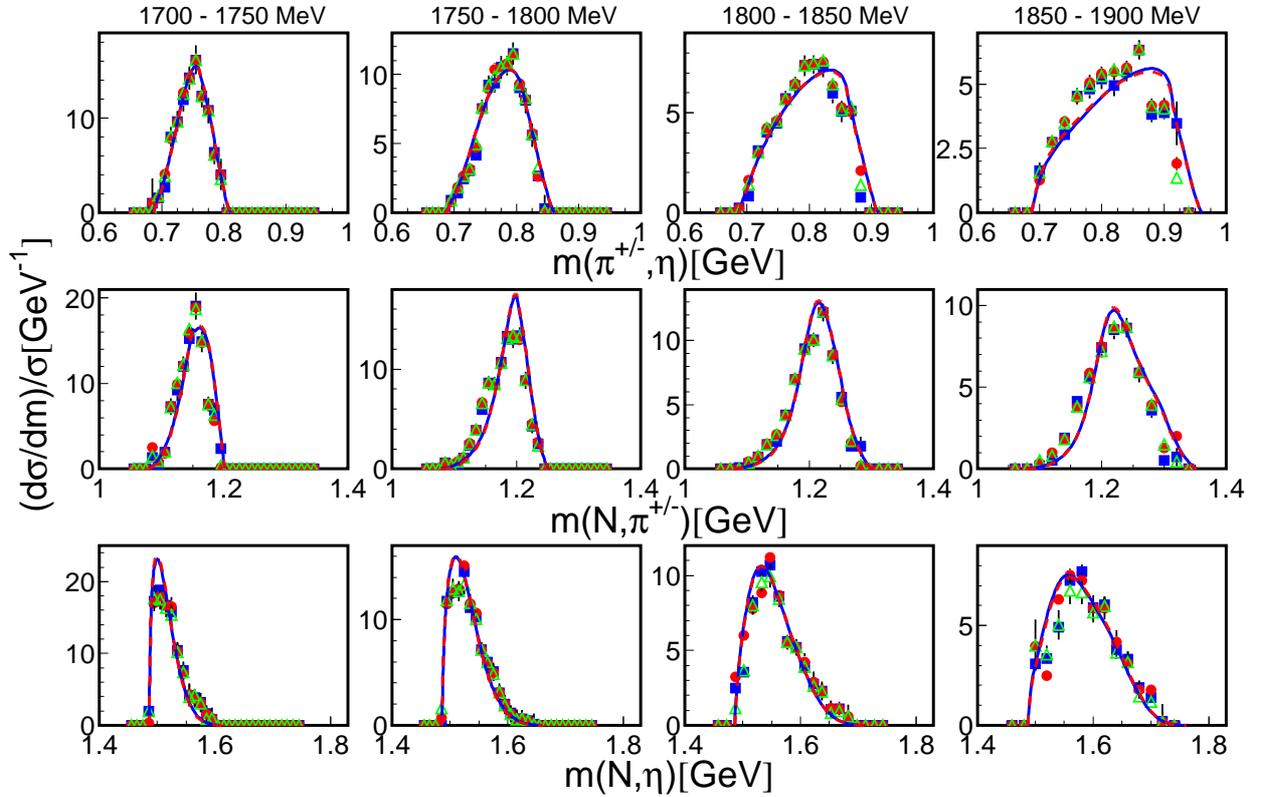}
}}
\caption{Invariant mass distributions for the $\pi^{\pm}\eta N$ final state. 
Upper row:  $\eta$ - pion invariant mass, central row: nucleon - pion, bottom row: nucleon - $\eta$.
Notation is the same as in Fig.~\ref{fig:theta_eta}. 
}
\label{fig:mpi0n}       
\end{figure*}

\clearpage

\begin{figure*}[thb]
\centerline{\resizebox{1.\textwidth}{!}{%
  \includegraphics{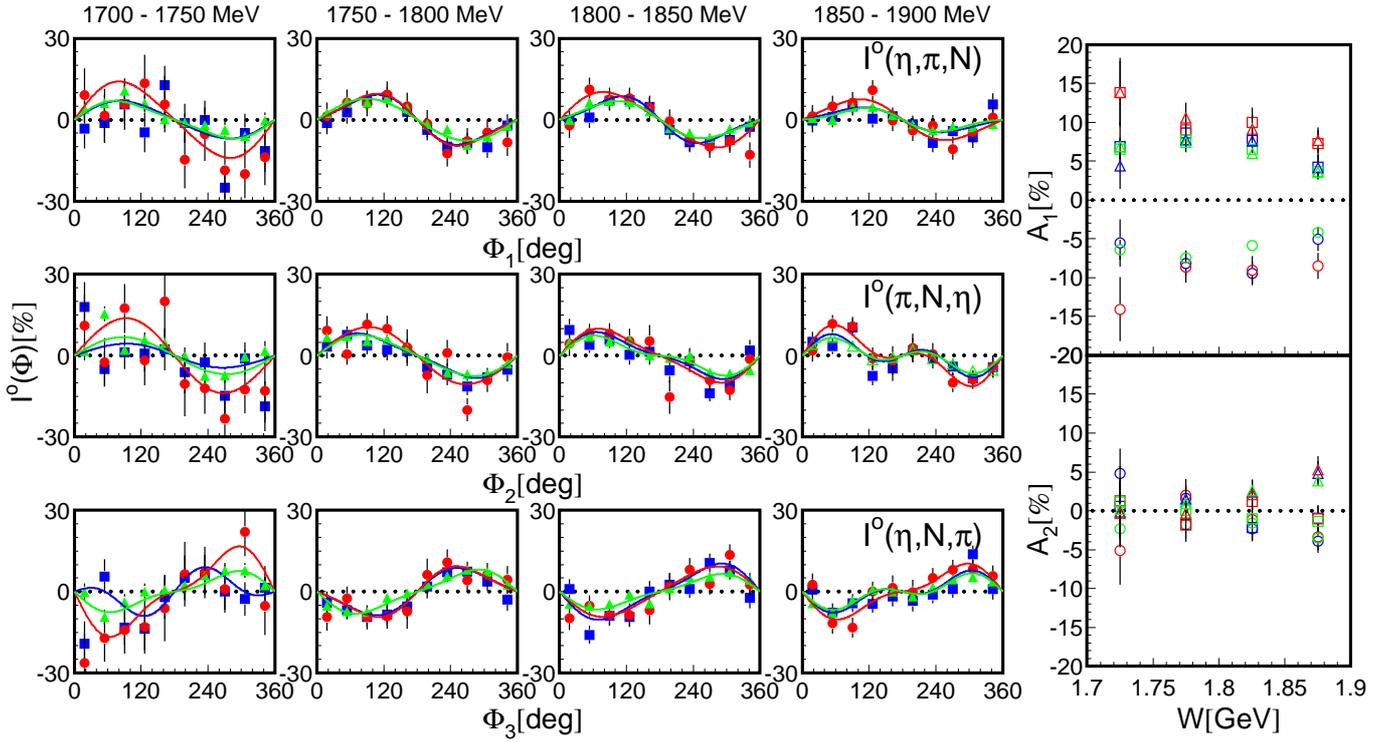}
}}
\caption{Left hand side: Beam-helicity asymmetries $I^{\odot}(\Phi)$ for the reactions $\gamma N\rightarrow N\pi^0\eta$
for different $W$ ranges given at top of Fig.
The three rows correspond to the asymmetries $I^{\odot}(\eta,\pi,N)$, $I^{\odot}(\pi,N,\eta)$,
and $I^{\odot}(\eta,N,\pi)$ defined in Sec.~\ref{sec:observables}. (Green) triangles: free protons
in initial state, (blue) squares: quasi-free protons, (red) circles: quasi-free neutrons. 
Solid curves: fits to data with Eq.~\ref{eq:coeff} (same color code as for data).
Panel at right hand side: fit coefficients $A_1$ and $A_2$ as defined in Eq.~(\ref{eq:coeff}). 
Colors indicate reaction type like above, 
open squares: $I^{\odot}(\eta,\pi,N)$, open triangles: $I^{\odot}(\pi,N,\eta)$, 
open circles: $I^{\odot}(\eta,N,\pi)$.
}
\label{fig:helipi0}       
\end{figure*}

\begin{figure*}[htb]
\centerline{\resizebox{1.\textwidth}{!}{%
  \includegraphics{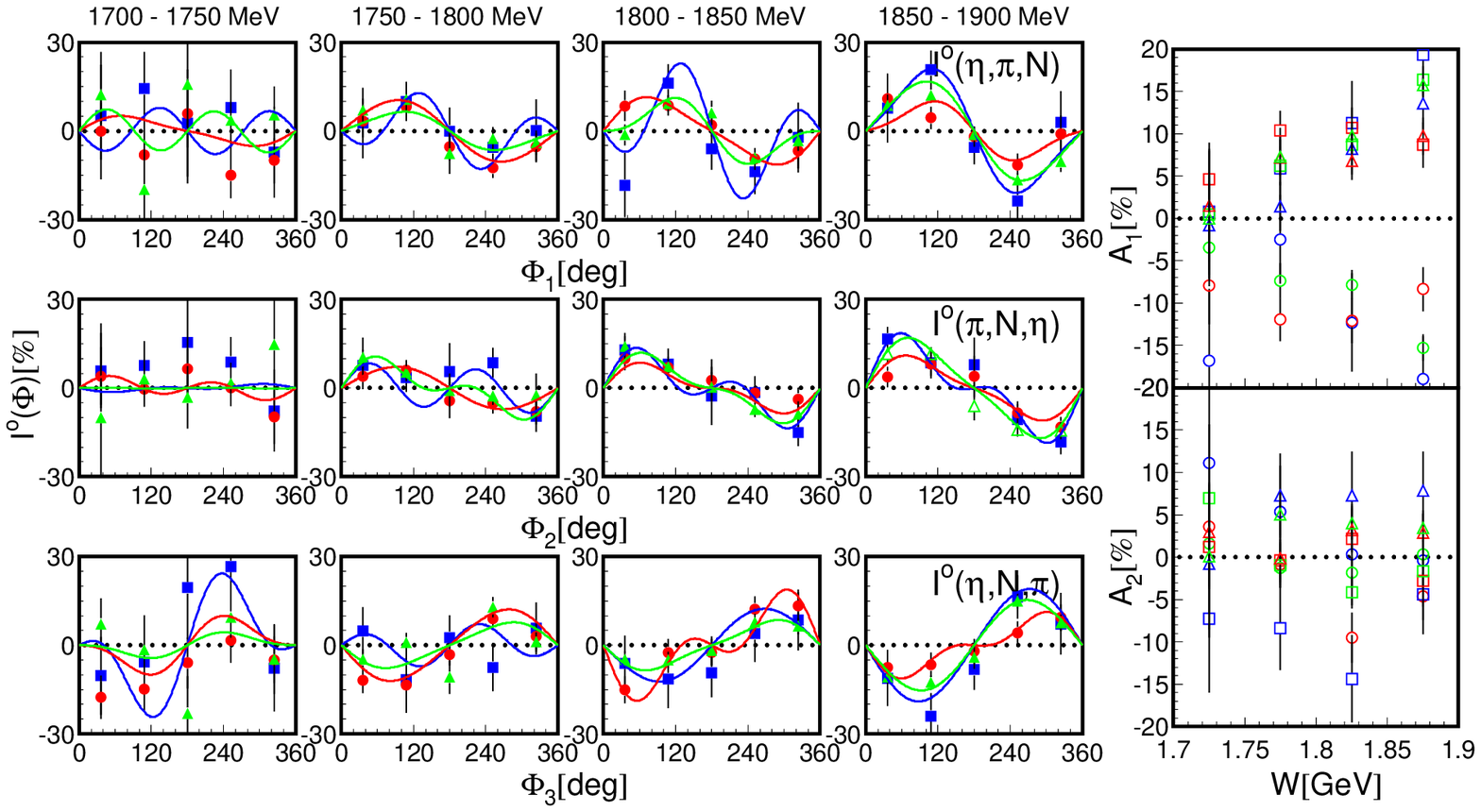}
}}
\caption{Same as Fig.~\ref{fig:helipi0} for charged pions.} 
\label{fig:helipic}       
\end{figure*}

\clearpage

\begin{figure*}[thb]
\centerline{\resizebox{1.\textwidth}{!}{%
  \includegraphics{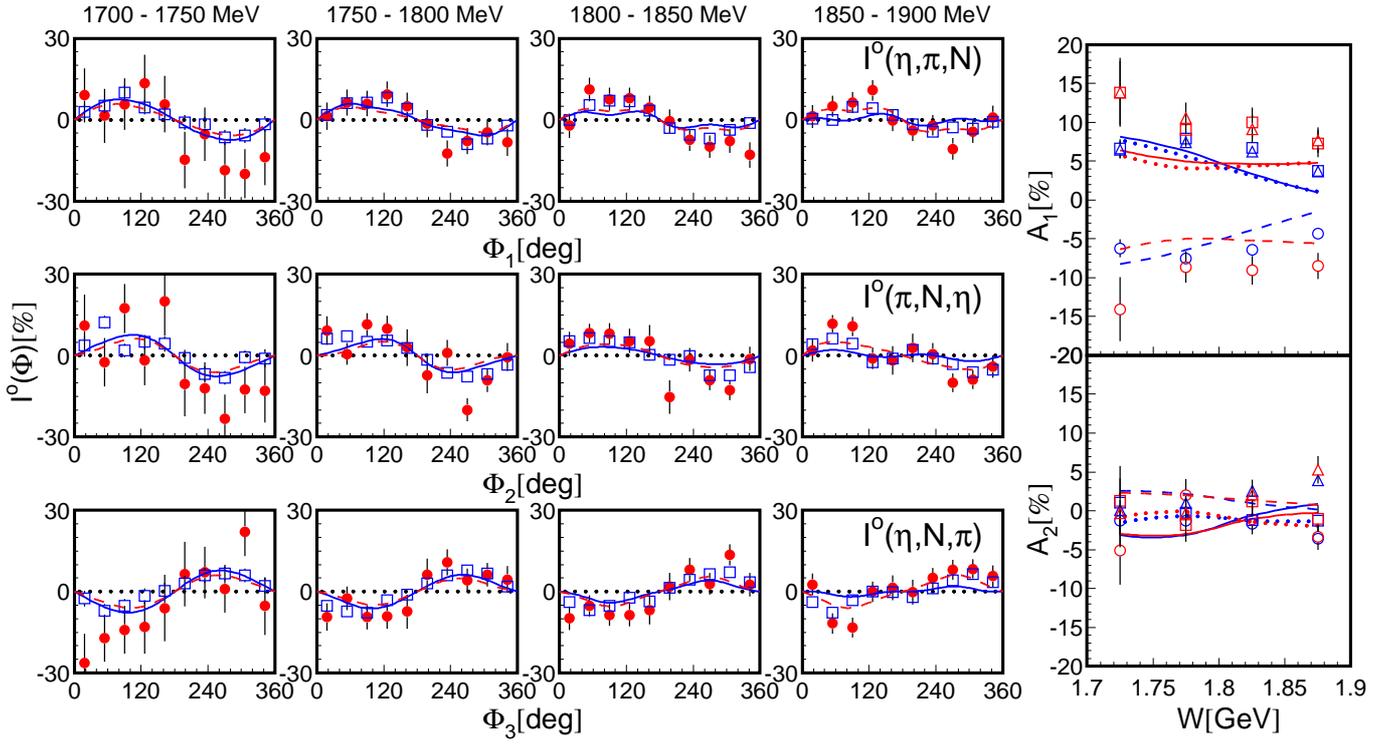}
}}
\caption{Beam-helicity asymmetries for the reactions $\gamma N\rightarrow N\pi^0\eta$
compared to predictions from Ref.~\cite{Fix_10}.  
(Blue) open squares: average of free and quasi-free proton data shown in Fig.~\ref{fig:helipi0}, 
(red) circles: quasi-free neutrons (same data as in Fig.~\ref{fig:helipi0}). Curves: model predictions from \cite{Fix_10},
solid (blue): proton target, dashed (red) neutron target.
Panel at right hand side: fit coefficients $A_1$ and $A_2$. 
 Colors indicate reaction type like above, open squares:
$I^{\odot}(\eta,\pi,N)$, open triangles: $I^{\odot}(\pi,N,\eta)$, open circles: $I^{\odot}(\eta,N,\pi)$.
Curves: model predictions from \cite{Fix_10}; blue (red) for proton (neutron) target;  
solid, dashed, dotted for $I^{\odot}(\eta.\pi^0,p,)$, $I^{\odot}(\eta.N,\pi)$, $I^{\odot}(\pi,N,\eta)$.
}
\label{fig:helipi0f}       
\end{figure*}

\begin{figure*}[htb]
\centerline{\resizebox{1.\textwidth}{!}{%
  \includegraphics{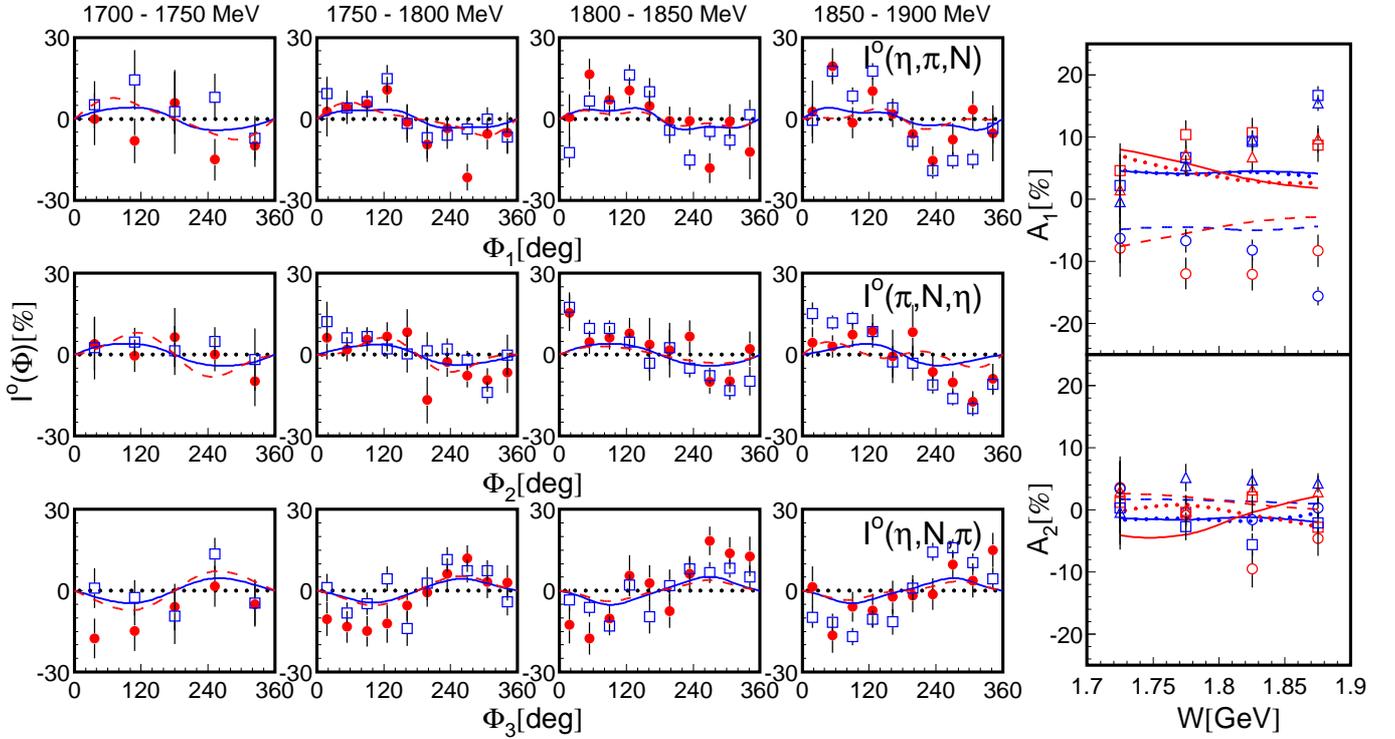}
}}
\caption{Same as Fig.~\ref{fig:helipi0f} for charged pions.} 
\label{fig:helipicf}       
\end{figure*}

\clearpage

Angular distributions of the pions in the canonical and helicity frames (see Sec.~\ref{sec:observables}
for definitions) for the reactions with neutral pions are summarized in Fig.~\ref{fig:ang_pi0} and for 
charged pions in Fig.~\ref{fig:ang_pic}. Similar observations as for the $\eta$ polar-angle distributions 
can be made. In general, the free and quasi-free proton data are quite similar, indicating only small FSI 
effects on the shape of the angular distributions. The measured angular distributions show only small 
variations between different target nucleons and the charge type of the final state pions and are in 
overall fair agreement with the model predictions. 

Invariant mass distributions of the meson-meson and meson-nucleon pairs are shown in 
Figs.~\ref{fig:mpi0p} and \ref{fig:mpi0n}. These data are in almost perfect agreement for free and
quasi-free protons and quasi-free neutrons as targets and for final state neutral and charged pions.
They agree with the model predictions from \cite{Fix_10} which are also essentially identical for all 
reaction channels. 

Some deviations between experimental data and predictions occur in the peaks of the narrow 
distributions, but this may be caused partly by experimental resolution effects, which 
have not been folded into the model predictions. The shapes of the invariant mass distributions
are dominated by the sequential $\Delta^{\star}\rightarrow\Delta\eta\rightarrow N\eta\pi$ decay chain.
All reactions show a pronounced peak in the $N\pi$ invariant mass from the intermediate $\Delta$(1232)
state (for the lowest photon energies the intermediate $\Delta(1232)$ can be only
populated at invariant masses below its peak value). There are no particular structures from the 
$N(1535)1/2^-\pi$ intermediate state visible in the $N\eta$ invariant mass distributions.
In previous measurements of the $\gamma p\rightarrow p\pi^0\eta$ reaction \cite{Gutz_14}, 
such contributions only became prominent at higher photon energies. They are expected close 
to the lower phase-space limit of the distributions where they are difficult to identify (see, however, 
the discussion below of the asymmetries which show more evidence for such contributions). The $\pi\eta$ 
invariant masses are also structureless because the $a_0$ meson with its strong decay to $\pi\eta$ 
has a mass of 980 MeV and lies thus outside the range of the current measurements.
 
Finally, the beam-helicity asymmetries introduced in Sec.~\ref{sec:observables} have been analyzed.  
The results are shown in Figs.~\ref{fig:helipi0} and \ref{fig:helipic}. They have been fitted with
the expansion from Eq.~\ref{eq:coeff} and the fit results for the expansion coefficients are also shown. 
This type of asymmetry is usually very sensitive to small reaction amplitudes such as 
from background terms (see e.g. \cite{Krambrich_09,Oberle_13,Oberle_14} for similar results for pion pairs). 
Again, the comparison of free and quasi-free proton results indicates no significant FSI effects on this 
observable. The results for target protons and neutrons are similar for all three asymmetries 
over the entire investigated energy range. The comparison of the reactions with neutral and charged pions 
is not so stringent due to the rather poor statistical quality of the data for charged pions, but 
qualitatively, those data are also quite similar. As already reported in \cite{Kashevarov_10} for
$I^{\odot}(\pi^0,p,\eta)$, only the $A_1$ coefficient is significantly different from zero. 
The $A_1$ magnitude is similar for reactions with charged and neutral pions. There are suggestions in 
the data of small differences in the energy dependence of the beam helicity asymmetries for charged 
and neutral pions, although the statistical uncertainties are large, particularly for charged pions 
at low energies.

The results are compared in Figs.~\ref{fig:helipi0f} and \ref{fig:helipicf} to the predictions from
Ref.~\cite{Fix_10}. One should note that these model results are really predictions, as previously 
only a few results for $I^{\odot}(\pi^0,p,\eta)$ reported in \cite{Kashevarov_10} were available. 
Data for the other isospin channels and the other types of asymmetry have been measured here for the 
first time and were not included in the model fits. For simplification of the figures and for better
statistical quality, the experimental results for free and quasi-free proton targets have been 
averaged (the comparison in Figs.~\ref{fig:helipi0} and \ref{fig:helipic} did not reveal any systematic
differences). The agreement between experimental data and model predictions is quite good. As long as the
$3/2^-$ wave dominates, the $A_1$ coefficient comes from the interference of the $\eta\Delta(1232)$ and 
$\pi N(1535)$ decay modes of the two $3/2^-$ resonances $\Delta(1700)$ and $\Delta(1940)$. 
Therefore, the observed behavior is evidence for contributions from the $\pi N(1535)$ intermediate
state, which is difficult to establish in the invariant mass distributions. 
Other resonances and background mechanisms mostly contribute to higher order terms in Eq.~\ref{eq:coeff} 
(which are almost insignificant in the experimental results within the achieved statistical accuracy).
In particular, as discussed in \cite{Kashevarov_10}, the coefficient $A_2$ is due to interference of 
the positive parity states with the dominant $3/2^-$ wave.  

\section{Summary and Conclusions}

Photoproduction of $\pi\eta$ pairs off nucleons has been studied for all possible isospin channels using
a liquid deuterium target (i.e. with quasi-free nucleons bound in the deuteron) and, for comparison, for
the reactions $\gamma p\rightarrow p\eta\pi^0$ and $\gamma p\rightarrow n\eta\pi^+$, with a free
proton target (liquid hydrogen). For all reactions, total cross sections, various angular distributions,
invariant mass distributions of meson-meson and meson-nucleon pairs, and all possible types of
beam-helicity asymmetries (circularly polarized photon beam) have been measured. For the quasi-free
measurements, the effects from Fermi motion have been eliminated by a full kinematic reconstruction
of the final state.

The major findings are the following: the absolute scale of the cross sections for free and quasi-free
protons is different. Total cross sections for production of $\pi^0\eta$ pairs off protons bound in the 
deuteron are suppressed with respect to the free proton to roughly 75\%. For protons bound in $^3$He nuclei,
the reduction is to 60\%. This is a clear indication of FSI effects, which are larger for the more strongly
bound helium nucleus. As in the case for reactions with single pion production, FSI effects are smaller for
final states with charged pions (reduced to 90\% for protons bound in the deuteron). This  
is most likely due to the different FSI in the $nn$ and $pp$ systems compared to the $np$ system, but
there is so far no quantitative modeling of these FSI effects. The measured asymmetries show no significant 
differences between free and quasi-free protons and also the angular and invariant-mass distributions
agree quite well after renormalization to the scale of the total cross sections. Thus one can conclude that
the FSI effects influence mainly the absolute scale of the cross sections, but are not so important for
polarization observables and for shapes of differential cross sections (the same observation has previously
been made for photoproduction of pion pairs \cite{Oberle_13,Oberle_14,Dieterle_15}).

The ratios of the total cross sections for the different isospin channels are all in good agreement
with Eq.~\ref{eq:isorel} which was derived under the assumption of a dominant
$\gamma N\rightarrow\Delta^{\star}\rightarrow \Delta(1232)\eta \rightarrow N\eta\pi$ reaction chain.
Apart from the FSI related effects noted in the absolute cross sections, all differential cross sections 
agree quite well
with a simple isobar model \cite{Fix_10} based on a dominant contribution from the above reaction 
with $\Delta^{\star}$ = $\Delta 3/2^-$ ($\Delta(1700)$ at threshold, $\Delta(1940)$ at higher energies).
The beam-helicity asymmetries are naturally explained in the framework of this model when interferences
between the decays of the $\Delta^{\star}3/2^-$ states to the $\eta\Delta(1232)3/2^+$ and $\pi N(1535)1/2^-$ 
final states are considered. 
In addition, small contributions from a few further $\Delta$ resonances (which manifest themselves mainly 
in angular distributions), very minor contributions from nucleon Born terms, and no contributions at all 
from isospin $I=1/2$ $N^{\star}$ resonances are required in the model. The latter would destroy the simple 
Clebsch-Gordan coefficient relation from Eq.~\ref{eq:isorel} due to the different photo-couplings of 
$N^{\star}$ resonances for protons and neutrons. This special situation has made it possible to make good 
model predictions for all isospin channels based on experimental results for the 
$\gamma p\rightarrow p\eta\pi^0$ reaction. This is very different from single $\eta$ or single $\pi$ 
photoproduction. In the latter case abundant results for different observables for three of the four possible 
isospin channels were available ($\gamma p\rightarrow p\pi^0$, $\gamma p\rightarrow n\pi^+$, 
$\gamma n\rightarrow p\pi^-$). Since there are only three independent isospin amplitudes, (as in Eq.~\ref{eq:iso})
this should be enough to predict the results for the fourth channel ($\gamma n\rightarrow n\pi^0$). 
However, results from different models do not agree and none of them predicted correctly the recently reported 
experimental data \cite{Dieterle_14} for the fourth reaction.

Photoproduction of $\pi\eta$ pairs is an efficient tool for the study of $\Delta$ excitations at moderate 
energies, in particular for the $\Delta(1700)3/2^-$. This reaction will most likely become a similar 
benchmark for this state as single $\eta$ production is for the $N(1535)1/2^-$ \cite{Krusche_95,Krusche_97}.
It seems that already now at the current energies, production of $\eta\pi$ pairs is much better understood than 
production of pion pairs, for which results from different models vary strongly and are not in good 
agreement with experimental data \cite{Dieterle_15}.  

\vspace*{1cm}
{\bf Acknowledgments}

We wish to acknowledge the outstanding support of the accelerator group and operators of MAMI. 
This work was supported by Schweizerischer Nationalfonds (200020-156983, 132799, 121781, 117601, 113511), 
Deutsche For\-schungs\-ge\-mein\-schaft (SFB 443), the INFN-Italy, the European Community-Research 
Infrastructure Activity under FP7 programme (Hadron Physics2, grant agreement No. 227431), 
the UK Science and Technology Facilities Council (ST/J000175/1, ST/G008604/1, ST/G008582/1, ST/J00006X/1), 
the Natural Sciences and Engineering Research Council (NSERC, FRN: SAPPJ-2015-00023), \newline Canada. This material 
is based upon work also supported by the U.S. Department of Energy, Office of Science, Office of Nuclear 
Physics Research Division, under Award Numbers DE-FG02-99-ER41110, DE-FG02-88ER40415, and \newline DE-FG02-01-ER41194 
and by the National Science Foundation, under Grant Nos. PHY-1039130 and IIA-1358175.
A. Fix acknowledges support from the MSE program Nauka (Project 3.825.2014/K).
We thank the undergraduate students of Mount Allison University and The George Washington University 
for their assistance.


\begin{thebibliography}{99}
\bibitem{Chiang_97}             W.T. Chiang and F. Tabakin,                       Phys. Rev.                      C {\bf  55},         2054  (1997).
\bibitem{Roberts_05}            W. Roberts and T. Oed,                            Phys. Rev.                      C {\bf  71},       055201  (2005).
\bibitem{Sarantsev_08}          A.V. Sarantsev {\it et al.},                      Phys. Lett.                     B {\bf 659},           94  (2008).
\bibitem{Thoma_08}              U. Thoma {\it et al.},                            Phys. Lett.                     B {\bf 659},           87  (2008).
\bibitem{Krambrich_09}          D. Krambrich {\it et al.},                        Phys. Rev. Lett.		    {\bf 103},       052002  (2009).
\bibitem{Kashevarov_12}         V.L. Kashevarov {\it et al.},                     Phys. Rev.			  C {\bf  85},       064610  (2012).
\bibitem{Zehr_12}               F. Zehr {\it et al.},                             Eur. Phys. J. 		  A {\bf  48},  	 98  (2012).
\bibitem{Oberle_13}             M. Oberle {\it et al.},                           Phys. Lett.			  B {\bf 721},  	237  (2013).
\bibitem{Oberle_14}             M. Oberle {\it et al.},                           Eur. Phys. J. 		  A {\bf  50},  	 54  (2014).
\bibitem{Thiel_15}              A. Thiel {\it et al.},                            Phys. Rev. Lett.		    {\bf 114},       091803  (2015).
\bibitem{Sokhoyan_15a}          V. Sokhoyan {\it et al.},                         Phys. Lett.			  B {\bf 746},  	127  (2015).
\bibitem{Sokhoyan_15b}          V. Sokhoyan {\it et al.},                         Eur. Phys. J. 		  A {\bf  51},  	 95  (2015).
\bibitem{Dieterle_15}           M. Dieterle {\it et al.},                         Eur. Phys. J.                   A {\bf  51},          142  (2015).
\bibitem{Nakabayashi_06}        T. Nakabayashi {\it et al.},                      Phys. Rev.			  C {\bf  74},       035202  (2006).
\bibitem{Ajaka_08}              J. Ajaka {\it et al.},                            Phys. Rev. Lett.		    {\bf 100},       052003  (2008).
\bibitem{Horn_08a}              I. Horn {\it et al.},                             Phys. Rev. Lett.		    {\bf 101},       202002  (2008).
\bibitem{Horn_08b}              I. Horn {\it et al.},                             Eur. Phys. J. 		  A {\bf  38},  	173  (2008).
\bibitem{Gutz_08}               E. Gutz {\it et al.},                             Eur. Phys. J. 		  A {\bf  35},  	291  (2008).
\bibitem{Gutz_10}               E. Gutz {\it et al.},                             Phys. Lett.			  B {\bf 687},  	 11  (2010).
\bibitem{Gutz_14}               E. Gutz {\it et al.},                             Eur. Phys. J. 		  A {\bf  50},  	 74  (2014).
\bibitem{Kashevarov_09}         V. Kashevarov {\it et al.},                       Eur. Phys. J. 		  A {\bf  42},  	141  (2009).
\bibitem{Kashevarov_10}         V. Kashevarov {\it et al.},                       Phys. Lett.			  B {\bf 693},  	551  (2010).	 
\bibitem{Fix_10}                A. Fix, V.L. Kashevarov, A. Lee, and M. Ostrick,  Phys. Rev.                      C {\bf  82},       035207  (2010), and priv. com.
\bibitem{Fix_11}                A. Fix and H. Arenh\"{o}vel,                      Phys. Rev.                      C {\bf  83},       015503  (2011). 
\bibitem{Krusche_15}            B. Krusche and C. Wilkin,                         Prog. Part. Nucl. Phys.           {\bf  80},           43  (2015).
\bibitem{Doering_06a}           M. D\"oring, E. Oset, D. Strottman,               Phys. Lett.                     B {\bf 639},           59  (2006).
\bibitem{Doering_06b}           M. D\"oring, E. Oset, D. Strottman,               Phys. Rev.                      C {\bf  73},       045209  (2006).
\bibitem{Kaeser_15}             A. K\"aser {\it et al.},                          Phys. Lett.                     B {\bf 748},          244  (2015).
\bibitem{Krusche_11}            B. Krusche,                                       Eur. Phys. J. Special Topics      {\bf 198},          199  (2011).
\bibitem{Werthmueller_13}       D. Werthm\"{u}ller {\it et al.},                  Phys. Rev. Lett.		    {\bf 111},       232001  (2013).
\bibitem{Werthmueller_14}       D. Werthm\"{u}ller {\it et al.},                  Phys. Rev.			  C {\bf  90},       015205  (2014).
\bibitem{Dieterle_14}           M. Dieterle {\it et al.},                         Phys. Rev. Lett.		    {\bf 112},       142001  (2014).	
\bibitem{Pheron_12}             F. Pheron {\it et al.},                           Phys. Lett.                     B {\bf 709},           21  (2012).       
\bibitem{Witthauer_13}          L. Witthauer {\it et al.},                        Eur. Phys. J.                   A {\bf  49},          154  (2013).
\bibitem{Herminghaus_83}        H. Herminghaus {\it et al.},                      IEEE Trans. on Nucl. Science.     {\bf  30},         3274  (1983).
\bibitem{Walcher_90}            Th. Walcher,                                      Prog. Part. Nucl. Phys.           {\bf  24},          189  (1990).
\bibitem{Anthony_91}            I. Anthony {\it et al.},                          Nucl. Inst. and Meth.           A {\bf 301},          230  (1991).
\bibitem{Hall_96}               S.J. Hall, G.J. Miller, R. Beck, P.Jennewein,     Nucl. Inst. and Meth.           A {\bf 368},          698  (1996).
\bibitem{McGeorge_08}           J.C. McGeorge {\it et al.},                       Eur. Phys. J.                   A {\bf  37},          129  (2008).
\bibitem{Olsen_59}              H. Olsen and L.C. Maximon,                        Phys. Rev.                        {\bf 114},          887  (1959). 
\bibitem{Novotny_91}            R. Novotny,                                       IEEE Trans. Nucl. Sci.            {\bf  38},          379  (1991).
\bibitem{Gabler_94}             A.R. Gabler {\it et al.},                         Nucl. Inst. and Meth.           A {\bf 346},          168  (1994).
\bibitem{Starostin_01}          A. Starostin et al.,                              Phys. Rev.                      C {\bf  64},       055205  (2001).
\bibitem{Watts_04}              D. Watts, in {\em Calorimetry in Particle Physics, Proceedings of the 11th Internatinal Conference, Perugia,
                                Italy 2004}, edited by C. Cecchi, P. Cenci, P. Lubrano, and M. Pepe (World Scientific, Singapore, 2005, p. 560). 
\bibitem{GEANT4}                S. Agostinelli et al.,                            Nucl. Instr. Meth.              A {\bf 506},          250  (2003).
\bibitem{Egorov_13}             M. Egorov and A. Fix,                             Phys. Rev.                      C {\bf  88},       054611  (2013).
\bibitem{Krusche_03}            B. Krusche and S. Schadmand,                      Prog. Part. Nucl. Phys.           {\bf  51},          399  (2003).
\bibitem{Pfeiffer_04}           M. Pfeiffer et al.,                               Phys. Rev. Lett.                  {\bf  92},       252001  (2004).
\bibitem{Krusche_95}            B. Krusche et al.,                                Phys. Rev. Lett.                  {\bf  74},       3736    (1995).
\bibitem{Krusche_97}            B. Krusche et al.,                                Phys. Lett. B                     {\bf 397},        171    (1997).


\end{thebibliography}
\end{document}